\documentclass[a4paper,11pt]{article}
\pdfoutput=1 % if your are submitting a pdflatex (i.e. if you have
\usepackage{jheppub} % for details on the use of the package, please
\usepackage[all]{xy}
\usepackage{doi}
\DeclareMathOperator{\sech}{sech}

\DeclareMathOperator{\arcsinh}{arcsinh}

\newcommand{\R}{\mathbb{R}}

\usepackage[mathlines]{lineno}% Enable numbering of text and display math
\modulolinenumbers[5]% Line numbers with a gap of 5 lines
\title{\boldmath Exact Ground States and Domain Walls in One Dimensional Chiral Magnets}

\author[a,b,1]{Calum Ross,\note{Corresponding author.}}
\author[a]{Norisuke Sakai,}
\author[a]{Muneto Nitta,}

\affiliation[a]{Department of Physics and Research and Education Center for Natural Sciences, Keio University,\\ Hiyoshi 4-1-1, Yokohama, Kanagawa 223-8521, Japan}
\affiliation[b]{Department of Mathematics, University College London, Gower Street, London WC1E 6BT, United Kingdom}

\emailAdd{c.ross@keio.jp}
\emailAdd{norisuke.sakai@gmail.com}
\emailAdd{nitta@phys-h.keio.ac.jp}

\abstract{We determine exactly the phase structure of a chiral magnet 
in one spatial dimension with the Dzyaloshinskii-Moriya (DM) 
interaction and a potential that is a function of the third component 
of the magnetization vector, $n_3$, with a Zeeman (linear with the 
coefficient $B$) term and an anisotropy (quadratic with the 
coefficient $A$) term, constrained so that $2A\leq \vert B\vert$. 
For large values of potential parameters $A$ and $B$, the system 
is in one of the ferromagnetic phases, whereas it is in the spiral phase 
for small values. 
In the spiral phase we find a continuum of spiral solutions, 
which are one-dimensionally modulated solutions with various 
periods. 
The ground state is determined as the spiral solution with the 
lowest average energy density. 
As the phase boundary approaches, the period of the lowest 
energy spiral solution diverges, and the spiral solutions become domain wall solutions 
with zero energy at the boundary. 
The energy of the domain wall solutions is positive in the homogeneous 
phase region, but is negative in the spiral phase region, signaling 
the instability of the homogeneous (ferromagnetic) state. 
The order of the phase transition between spiral and homogeneous 
phases and between polarized ($n_3=\pm 1$) and canted ($n_3\not=\pm 1$) 
ferromagnetic phases is found to be second order.}

\arxivnumber{2012.08800}

\begin{document} 
\maketitle
\flushbottom

\section{Introduction}
\label{introduction}
Chiral magnets are special examples of magnetic materials where 
the energy functional which describes the system contains a term 
with a preferred chirality for the magnetization vector 
\cite{NT,YOKPHMNT,MBJPRNGB,RHMBWVKW}. 
This chirality preference comes from the parity (inversion) violating 
(noncentrosymmetric) interaction called the Dzyaloshinskii-Moriya 
(DM) interaction consisting 
of an inner product of the magnetization vector $\vec{n}$ 
with the curl $\nabla \times \vec{n}$ of the
magnetization vector \cite{Dzyaloshinskii,Moriya}. 
The chiral magnet models are considered in various 
spatial dimensions with a variety of potentials for the 
magnetization vector besides the DM interaction term and the 
exchange term (square of derivative of magnetization vector). 
The ground state of the system is a translationally invariant 
homogeneous configuration, if the DM interaction is weak compared 
to the potential for the magnetization vector. 
If the DM interaction becomes more important than the potential, 
however, spatially modulated solutions become more favorable, 
and can become the ground state, which is spatially inhomogeneous 
and breaks the translation invariance. 

The spiral phase is one such interesting phase, with
various terminologies used for various specific cases 
in the existing literature for magnetic systems. 
One of the first and simplest examples is the case 
of no potential (with only exchange term and a DM interaction), 
which is called a chiral helimagnet. 
When the potential consists only of a Zeeman term (a 
term proportional to the component $n_3$ of the magnetization vector), 
the spiral solution is called a chiral soliton lattice. 
These specific spiral solutions are discussed in detail in~\cite{KO2015,TOPSK_2018}. 
Both of these terms refer to spiral solutions in a chiral 
magnet without an anisotropy term (quadratic in the component $n_3$ 
of the magnetization vector) in the potential. 
In this work, we will use the terminology of spiral solutions 
for generic cases\cite{LSB}, and use the term chiral soliton lattice 
for the specific case of no anisotropy, although
chiral soliton lattice is sometimes used for generic spiral 
solutions~\cite{MRG2016}.

In high energy physics, a similar situation occurs, 
with chiral soliton lattice states possible due to both magnetic 
effects~\cite{Brauner:2016pko} and vortical effects~\cite{Nishimura:2020odq}. 
Another related one-dimensional modulated state is the nematic 
phase of chiral nematic liquid crystals \cite{deGennes,Dierking,Collings}. 
The Frank free energy which describes liquid crystals is known 
to become equivalent to the free energy of a chiral magnet in a 
suitable limit~\cite{FKATDS} and skyrmion configurations can be 
observed experimentally in liquid crystals. 
This correspondence is also commented on in~\cite{KO2015}. 
Another interesting inhomogeneous phase is the skyrmion lattice 
phase~\cite{LSB}.  
Skyrmion solutions were first studied in \cite{BY} in two dimensions.  
Among the spatially modulated solutions, magnetic skyrmions are 
intrinsically two dimensional \cite{BY,BH}.
In contrast, the spiral solutions are modulated along only one 
spatial dimension, and persist in the one dimensional system. 
In some special situations, spiral solutions have been constructed 
exactly~\cite{Han_2010, KO2015}. 
In the case of antiferromagnetic materials, various interaction 
terms due to the sublattice structure are considered phenomenologically, 
and various inhomogeneous solutions have been extensively 
studied and the phase diagram obtained~\cite{BRWM,BS1999}. 
For a different combinations of potential and the DM interaction 
term in one spatial dimension, instanton solutions have been 
exhaustively studied~\cite{Hongo:2019nfr}. 
Effective low-energy field theories in the spiral phase has 
been worked out to yield anisotropic dispersion 
relations\cite{Hongo:2020xaw}. 
When one considers chiral magnets in three spatial dimensions, a
richer phase structure emerges including modulated solutions 
in three spatial directions such as the cone or elliptic cone 
phases \cite{Rowland_2016}.

If we restrict ourselves to one spatial dimension, we obtain 
the one-dimensional chiral magnet.
It can offer a simpler system where we can study spiral solutions 
extensively. 
The field equations  
for a one dimensional model of a chiral 
magnet are closely related to the double sine-Gordon model. 
When there is no DM interaction, the model has been studied 
extensively, to obtain in particular domain wall (kink) solutions 
and their thermodynamic properties for the double sine-Gordon chain 
\cite{Condat1983}. 
The double sine-Gordon model also appears in high energy 
contexts such as two Higgs doublet models 
\cite{Eto:2018hhg,Eto:2018tnk} and $^3P_2$ neutron superfluids 
\cite{Chatterjee:2016gpm}, but without an interaction of the DM type.
One should note, however, that the presence of the DM interaction 
is crucially important to understand the energetics of spiral solutions 
and the phase structure of chiral magnets. 
The double sine-Gordon domain wall solutions have also been 
discussed in the context of two dimensional chiral magnets, 
in particular at their 
one-dimensional edges~\cite{MRG2016}.

The purpose of our paper is to introduce a model of a chiral 
magnet in one-spatial dimension to obtain the phase diagram 
exactly, and clarify its relation to energetics of domain wall solutions.
Alongside the usual energy term for the Heisenberg ferromagnet 
the model includes a Dzyaloshinskii-Moriya (DM) interaction term 
with the coefficient $\kappa$ \cite{Dzyaloshinskii,Moriya} and 
a potential which is a sum of a Zeeman term (linear in $n_3$) 
with the coefficient $B$ and an anisotropy term (quadratic in 
$n_{3}$) with the coefficient $A$. 
We find that the boundary energy functional needed to obtain the
Euler-Lagrange equation (through the variational principle) does 
not contribute to the solutions of interest to us here, such as spiral 
or domain wall solutions. 
We exhaustively work out the static solutions of the model, and 
find that spiral solutions have a continuous spectra of average 
energy density, and the lowest energy spiral solution gives the 
ground state in the spiral phase (for small potential parameters 
$A, B$). On the other hand, homogeneous (polarized or canted 
ferromagnetic) solutions give the ground state for large $A, B$. 
We demonstrate that the phase boundary between the spiral phase 
and the homogeneous (ferromagnetic) phases is characterized by 
the emergence of zero energy domain wall solutions. 
In the homogeneous (ferromagnetic) phase, the domain wall solution 
is a soliton as a finite positive energy excited state. 
In the spiral phase region, the domain wall solutions, 
as solitons above the homogeneous background, have negative 
energy signaling the instability of the homogeneous solution. 
The zero energy domain wall solutions are obtained as the 
(infinite period) limit of the spiral solutions when the phase 
boundary is approached from the spiral phase region. 
The exact phase boundary  in the $A/\kappa^2, B/\kappa^2$ plane with $2A\leq \vert B\vert$
is worked out explicitly. That is the transition between the spiral phase and the homogeneous, polarised, ferromagnetic phase which we show to be a second order transition.
%We find that the phase transition between the spiral phase and 
%the three distinct homogeneous phases are of second order. 
%Only the transition between positively ($n_3=1$) and negatively 
%($n_3=-1$) polarized ferromagnetic phases is of first order, 
%whereas the transition between polarized ($n_3=\pm 1$) and 
%canted ($n_3\not=\pm 1$) phases is of second order. 
We also obtain exact domain wall solutions explicitly 
for all parameter regions and confirm our general argument 
on the phase diagram. 

After finishing this work we became aware of the recent 
paper~\cite{Paterson}, as well as the older work \cite{IL1983}. 
The former obtains the phase boundary between homogeneous phases 
and the spiral phase using explicit spiral solutions in terms of 
elliptic functions, although they consider a different 
physical situation, namely with an external elastic strain applied 
to the chiral magnet. 
We have approached the problem from a complementary 
point of view, with general arguments using inequalities to show 
the region of the spiral phase. 
In particular, we have clarified the role of domain wall solutions 
explicitly to work out the phase boundary, and also have included 
more detail about the domain wall limit of the spiral solutions 
at the phase boundary. 
We have also explicitly derived the order of phase transition 
across the phase boundary. 

Another related work is \cite{Chovan} where a similar model, with just an anisotropy term and no magnetic field is considered. The model is studied numerically and two spiral phases are discussed; a ``flat" spiral similar to what we construct here, and a ``non-flat" spiral. The difference between the two types of spirals is apparent in a spherical coordinate decomposition of the magnetisation field. For flat spirals the angular variable $\Phi(x)$ is a constant, while for a non-flat spiral it is allowed to vary.  There is a numerical evidence that the non-flat spiral has lower energy than the flat spiral close to the transition to the homogeneous phase, and that it determines the location of the phase transition. The non-flat spiral has no analytic expression, here we focus on the flat spiral, just called spiral throughout, configurations which can be constructed explicitly. This is why we restrict to $2A\leq\vert B\vert$, as for $2A>\vert B\vert$ the non-flat spiral is the ground state and the phase transition cannot be found analytically. We still discuss details about spiral and domain wall solutions in this parameter region.

Our model can be obtained as a dimensional reduction of the most 
popular model of two-dimensional chiral 
magnets~\cite{BY,BH,BRS,Schroers1,RSN,DM,Melcher}. 
Hence one can expect that all the solutions that we consider 
are also solutions of  chiral magnet models in two 
or higher spatial dimensions. 
We would only need to additionally study other solutions specific to two spatial 
dimensions in order to determine the phase diagram of a two dimensional model. 
A similar problem was studied for the case of noncentrosymmetric 
uniaxial antiferromagnetic materials \cite{BRWM}, where the authors 
considered various interaction terms arising from sublattice 
magnetization vectors. 
As a result, physical consequences such as the  diagram are 
different, although the mathematical structure is similar.

The paper is organized as follows. 
In Sec.~\ref{sec:1dmodel} we introduce the one dimensional model, 
discuss its homogeneous phases and the vanishing boundary contributions. 
Sec.~\ref{sec:EOM} gives the field equations
and a classification 
of the different types of solutions. 
In Sec.~\ref{sec:spiral_solution}, the lowest energy spiral 
solution is shown to give the ground state in the spiral phase, and 
the phase boundary between the homogeneous phase is determined explicitly. 
The exact spiral solutions in terms of elliptic functions are 
also given for $B=0$ or $A=0$ cases. 
In Sec.~\ref{sec:exactdomainwalls} 
we present the exact domain wall solutions for the model. 
Sec.~\ref{sec:summary} contains a summary of the work 
presented in the paper and a discussion of some open questions. 
Appendix~\ref{sec:der_spiral} is devoted to the evaluation of the
integrals needed to obtain the second derivative of the average 
energy density of the spiral ground state near the phase boundary. 
Finally Appendix~\ref{sec:exact_sol_AorB} gives some details of 
exact spiral solutions for $B=0$ or $A=0$. 

\section{One dimensional chiral magnet}\label{sec:1dmodel}

\subsection{Model}\label{sec:field-eq}

Throughout this paper we study chiral magnets in one 
spatial dimension, in particular their phases and exact spiral 
and domain wall solutions. 
The energy density of the particular model we consider 
is given in terms of 
a real three component magnetization vector field 
$\vec{n}:\R^{1}\to S^{2}$ with the constraint $\vec{n}^{2}=1$ 
\begin{equation}
\mathcal{E}=\frac{1}{2}\left(\frac{ d\vec{n}}{dx}\right)^{2}
+\kappa\vec{n}\cdot \nabla_{-\alpha}\times \vec{n} +U(n_{3}) 
+\mathcal{E}_{\rm boundary}, 
\label{1D energy density}
\end{equation}
where the potential is
\begin{equation}
U(n_{3})=-Bn_{3}+An_{3}^{2}. 
\label{eq:potential term}
\end{equation}
We leave to the next subsection a discussion of the 
boundary term $\mathcal{E}_{\rm boundary}$, which does not 
affect the field equations but is needed to make the variational 
problem well-defined. 
The rotated gradient $\nabla_{-\alpha}$ is 
defined in terms of $\nabla=(\frac{d}{dx}, 0, 0)^T$ and a rotation 
matrix $R(-\alpha)$ by an angle $-\alpha\in (-2\pi, 0]$ about the 3-axis 
\begin{equation}
\nabla_{-\alpha}=R(-\alpha)\nabla=
\left(
\begin{array}{c}
\cos\alpha 
\\
-\sin\alpha
\\
0
\end{array}
\right) \frac{d}{dx}.
\label{eq:DMderivative}
\end{equation}
Using this the Dzyaloshinskii-Moriya (DM) interaction term 
\cite{Dzyaloshinskii,Moriya} can be written explicitly as 
\begin{equation}
\vec{n}\cdot \nabla_{-\alpha}\times \vec{n}
=\cos\alpha w_{B}+\sin\alpha w_{N},
\end{equation}
Here $w_{B}$ and $w_{N}$ are the Bloch and Ne\'{e}l DM terms which 
in one dimension become
\begin{equation}
w_{B}=n_{3}\frac{d n_{2}}{dx}-n_{2}\frac{d n_{3}}{dx}, 
\qquad w_{N}=n_{3}\frac{d n_{1}}{dx}-n_{1}\frac{d n_{3}}{dx},
\end{equation} 
respectively.

This model has the symmetry that the $B>0$ and $B<0$ parameter 
regions are related by sending $B\to -B$ and $n_{3}\to -n_{3}$. 
As such we chose to work with $B\geq 0$.

As observed in \cite{BRS} working in terms of a rotation by 
$\alpha$ around the $3$-axis, $R(\alpha)$, we can rewrite the DM term as
\begin{equation}
\vec{n}\cdot \nabla_{-\alpha}\times \vec{n}
=
\vec{n}^{\alpha}\cdot \nabla\times \vec{n}^{\alpha}, 
\end{equation}
with 
\begin{equation}
\vec{n}^{\alpha}=R(\alpha)\vec{n}. \label{rotated magnetisation vector}
\end{equation}
Since the first and third term in Eq.~\eqref{1D energy density} 
are invariant under $R(\alpha)$, Eq.~\eqref{1D energy density} 
can be rewritten in terms of $\vec{n}^{\alpha}$ and $\nabla$ 
replacing $\vec{n}$ and $\nabla_{-\alpha}$.

\subsection{Boundary terms}

A total derivative term in the action is a boundary 
term which does not affect the equations of motion. 
However, it can in general contribute to the energy similarly 
to the DM interaction term. 
It has been realized in Ref.~\cite{BRS} that there are subtle 
contributions from infinity (the boundary) to the energy of 
the skyrmion (hedgehog) solutions in the case of two-dimensional 
chiral magnets along the solvable line $2A=B$. 
Subsequently it has been realized that the additional total 
derivative term in the energy density is needed to make the 
variational principle well-defined, in deriving the field equations 
for the skyrmion \cite{RSN}. 
We call this term the boundary energy functional. 
In the two dimensional solvable case, the boundary energy functional gives a finite 
and crucial negative energy contribution to the skyrmion energy. 
This fact leads to the instability of the polarized ferromagnetic 
background solution and a phase transition to other phases, such as 
the skyrmion lattice phase, below the critical value of potential 
parameters \cite{RSN}. 

When one varies the energy functional to obtain the equation of 
motion, it requires a partial integration for the DM interaction 
term. 
The resulting surface term can be canceled by adding a boundary 
energy functional of the form \cite{RSN}
\begin{equation}
\mathcal{E}_{\rm boundary}=\kappa \epsilon_{ijk}\nabla_i
\left(n^{\alpha,{\rm bound}}_j n^{\alpha}_k\right), 
\label{eq:boundary-term}
\end{equation}
where $\vec{n}^{\alpha,{\rm bound}}$ is the boundary value of the 
magnetization vector. 
It is important to realize that the derivative is acting also on 
the boundary value $\vec{n}^{\alpha,{\rm bound}}(x)$, since 
it can depend on the position along the boundary, namely (two) 
isolated points in our one-dimensional model. 
Let us consider the total energy in an interval $-x_{1}\le x \le x_{2}$ 
with the boundaries located at $x=-x_1$ and $x=x_2$. 
The contribution from the boundary term is given by 
%\begin{widetext}
\begin{equation}
\begin{split}
E_{\rm boundary}
&=\kappa \int_{-x_1}^{x_2}dx\partial_1
\left(n_3^{{\rm bound}}(x)n^{\alpha}_2(x)
-n_2^{\alpha,{\rm bound}}(x)n_3(x)\right)\\
&=0, 
\end{split}
\label{eq:energy-boundary-term}
\end{equation}
%\end{widetext}
since the magnetization vector $\vec{n}^\alpha(x)$ takes the 
boundary value $\vec{n}^{\alpha, {\rm boundary}}(x)$  at the 
boundary. 
In the case of spiral solutions, we are interested in the total 
energy in one period, taking $x_2+x_1$ to be the period. 
In the case of domain wall solutions, we should choose $x_1, x_2\to \infty$. 
Thus we find no contribution from the boundary term for 
the class of solutions that we consider here. 
As such we will drop the boundary term from the energy density 
from now on.

\subsection{Homogeneous phases}

\begin{figure}[htbp]
\begin{center}\includegraphics[width=0.4\textwidth]{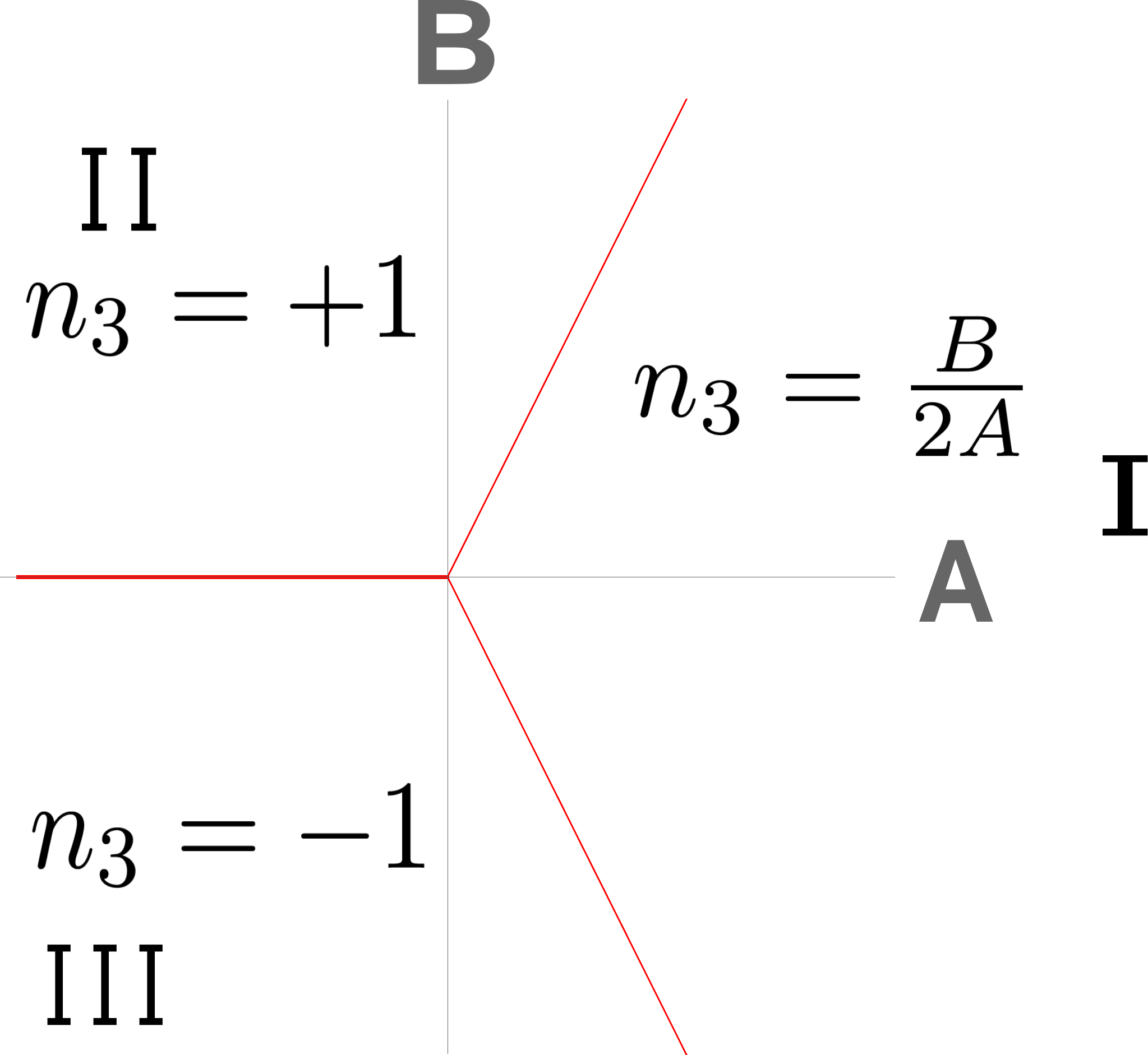}
\caption{The three ferromagnetic phases of the potential $U(n_{3})
=-Bn_{3}+An_{3}^{2}$.}
\label{phase fig}
\end{center}
\end{figure}
Our field equations admit not only one-dimensionally 
modulated solutions but also homogeneous configurations as 
solutions. 

They are the ground states in parameter regions with vanishing 
or small DM term.
Homogeneous solutions are given by the stationary points of 
the potential, and are in common with the model in 
two or higher spatial dimensions 
which has the same potential as in Eq.~\eqref{eq:potential term}, see for instance Refs.~\cite{BRS, RSN}.
 
Restricting to homogeneous solutions and 
comparing the energy density results in the familiar phase diagram shown 
in Fig.~\ref{phase fig}.

There are three distinct regions: 
\begin{enumerate}
\item
region I: $-2A\le B \le 2A$, with the minimum at $n_3=\frac{B}{2A}$, and 
\begin{equation}
{\cal E}_{\rm canted}=U_{\rm min}=-\frac{B^2}{4A}. 
\label{eq:min_canted}
\end{equation}
This is the canted ferromagnetic phase.
The minimum configuration of $\vec{n}$ is the 
circle $(n_{1})^{2}+ (n_{2})^{2}=1-\frac{B^{2}}{4A^{2}}$. 
%as shown in the middle of Fig.~\ref{vacuum manifold}. 
\item
region II: $B\ge 0$, and $B\ge 2A$, with minimum at $n_{3}= 1$, and 
\begin{equation}
{\cal E}_{+}=U_{\rm min}=A-B. 
\label{eq:min_pos_pol}
\end{equation}

This is the positively polarized ferromagnetic 
phase. 

%\begin{figure}
%\begin{center}\includegraphics[width=0.45\textwidth]{figures/vacuum_manifolds}
%\caption{The vacuum manifold showing how the ground state changes 
%depending on the relative size of $A$ and $B$. On the left when 
%$B>2A, B>0$ (region II) the ground state is $n_{3}=+1$, in 
%the middle with $-2A<B<2A$ (region I) the ground state is a circle 
%with $n_{1}^{2}+n_{2}^{2}=\frac{B}{2A}$, and on the right is a particular 
%case $B=0$ where the circle is the equator. \textcolor{red}{If the referee wants this section condensed this figure could easily be removed.}}
%\label{vacuum manifold}
%\end{center}
%\end{figure}

\item
region III: $B\le 0$, and $B\le 2A$, with the minimum of the potential 
\begin{equation}
{\cal E}_{-}=U_{\rm min}=A+B, 
\label{eq:min_neg_pol}
\end{equation}
at $n_{3}= -1$.
This is the negatively polarized ferromagnetic 
phase. 

\end{enumerate}

\begin{figure}[htbp]
\begin{center}
\includegraphics[width=0.45\textwidth]{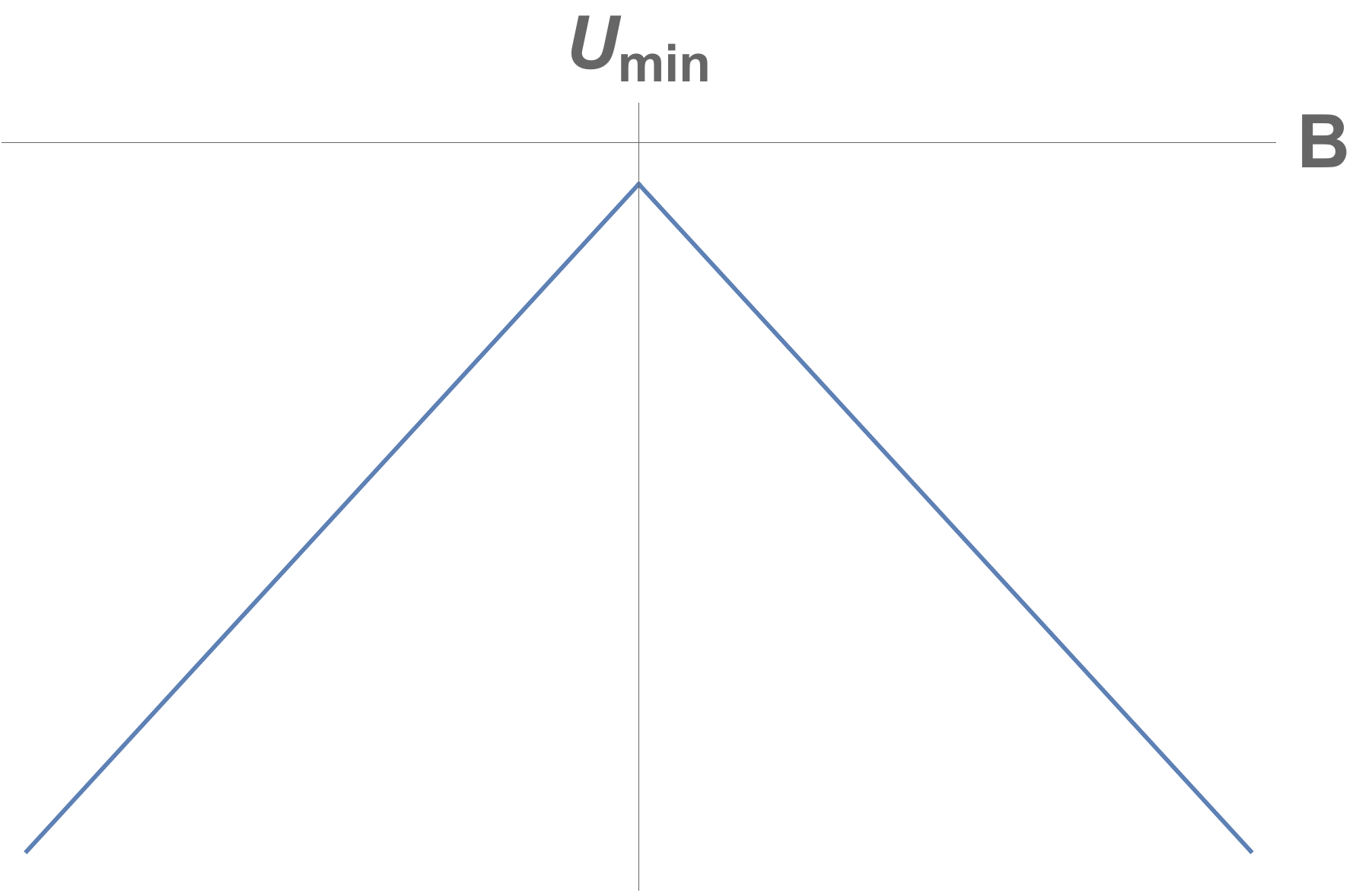}
\includegraphics[width=0.45\textwidth]{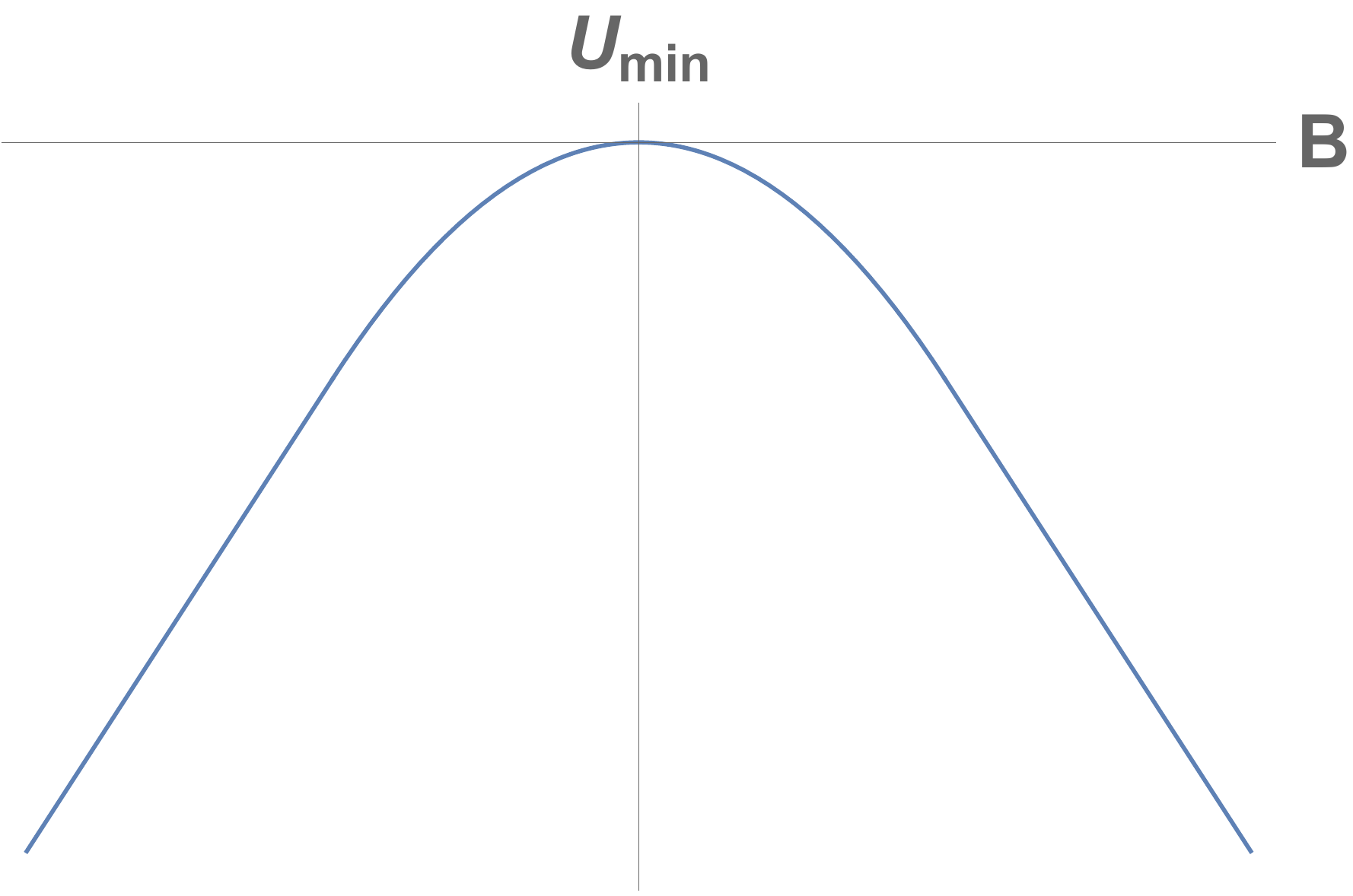}
\caption{The energy density $\mathcal{E}(A,B)=U_{\rm min}$ 
of the homogeneous ground state as a function of $B$ for fixed $A$. 
The left panel is for fixed $A<0$, exhibiting the first order 
phase transition between region II and III. 
The right panel is for $A>0$, exhibiting the second order phase 
transitions between regions II (positively polarized) and I 
(canted), and I and III (negatively polarized) phases. }
\label{fig:u_minvsB}
\end{center}
\end{figure}

The phase boundary lines are located at $B=2A\ge 0$ (between I and II), 
$B=0, A\le 0$ (between II and III), and $B=2A \le 0$ (between II and III).
%In this mean-field approximation, we can find out the order of 
%the phase transition between the three homogeneous phases by 
%simply comparing the energy density ${\mathcal E}_{\rm canted}$ in 
%\eqref{eq:min_canted}, ${\mathcal E}_{+}$ in \eqref{eq:min_pos_pol}, 
%and ${\mathcal E}_{-}$ in \eqref{eq:min_neg_pol} of the ground 
%states in the canted, positively polarized, and negatively 
%polarized ferromagnetic phases, respectively. 
The order of the phase transition was found previously 
\cite{Hongo:2019nfr}. 
\begin{figure}[htbp]
\begin{center}
\includegraphics[width=0.45\textwidth]{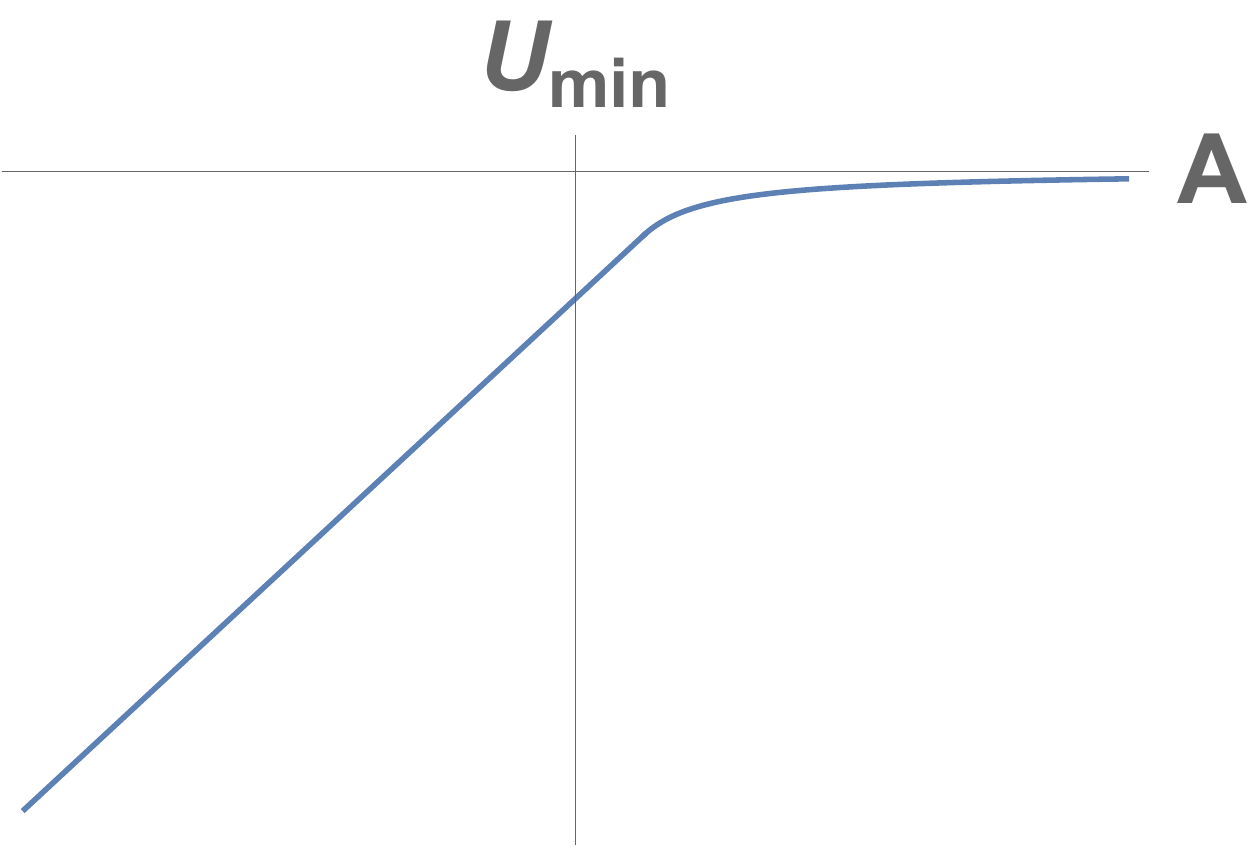}
\caption{The energy density $\mathcal{E}(A,B)=U_{\rm min}$ of the 
homogeneous ground state as a function of $A$ for a fixed $B\not=0$. 
It exhibit the second order transition between polarized 
($2A<|B|$) and the canted ($|B|<2A$) ferromagnetic phase. 
}
\label{fig:u_minvsA}
\end{center}
\end{figure}
The phase transition between regions I and II is of second order, since the energy density and its first 
derivative are continuous, while the second derivative 
is discontinuous. 
Similarly the phase transition between regions I and III
is also of second order. 
In Fig.~\ref{fig:u_minvsA}, we illustrate the second 
order phase transition between II and I ($B>0$) or III and I 
($B<0$). 
The phase transition between the regions II and III across the 
line $B=0, A<0$ is of first order, since the 
energy density is identical on both sides but its first 
derivative is discontinuous  
%$\frac{d{\mathcal E}_{ +}}{dB}\not=\frac{d{\mathcal E}_{-}}{dB}$
. 
Moreover, there are two distinct minima at $n_3=1$ and $n_3=-1$ 
with the same energy. 
In Fig.~\ref{fig:u_minvsB}, we illustrate the first 
order phase transition between II and III (for a fixed $A<0$) 
in 
the left panel and the second order phase transition between 
II and I, and one between I and III (for a fixed $A>0$) in the right panel. 

One can expect that homogeneous configurations tend to 
have lower energy when the potential term $U$ is more important 
than other terms, in particular the DM term. 
It is known that  spiral solutions 
become the ground states for parameter regions where the 
DM interaction is more important than the potential. 
We will determine the precise phase boundary.

%%%%%%%%%%
\section{The field equation}\label{sec:EOM}

\subsection{Conservation of ``mechanical energy'' }
\label{sec:conservation-law}

To find the solutions for the energy density in 
Eq.~\eqref{1D energy density}, 
it is easiest to first solve the constraint, $(\vec{n}^\alpha)^2=1$, 
using the independent angular variables $\Theta, \Phi$ 
\begin{equation}
\vec{n}^{\alpha}=(\sin\Theta\cos\Phi, \sin\Theta\sin\Phi, \cos\Theta)^T. 
\label{eq:angle-variable}
\end{equation}
The energy density in Eq.~\eqref{1D energy density} is given in 
terms of the independent fields $\Theta(x)$ and $\Phi(x)$ as 
%\begin{widetext}
\begin{equation}
\begin{split}
\mathcal{E}&=\frac{1}{2}\left\{\left(\frac{ d\Theta}{dx}\right)^{2}
+\left(\sin\Theta\frac{d\Phi}{dx}\right)^2\right\}+\kappa\left(\sin\Phi\frac{ d\Theta}{dx}
+\frac{1}{2}\sin 2\Theta \frac{d}{dx}\sin\Phi\right) 
+U(\Theta) ,
\end{split} 
\label{eq:energy-density-angle}
\end{equation}
%\end{widetext}
with the potential 
\begin{equation}
U(\Theta)=-B\cos\Theta+A\cos^{2}\Theta. 
\label{eq:potential-angle}
\end{equation}
Minimising the energy functional leads to the Euler-Lagrange or field equations for $\Theta$ 
and $\Phi$ being 
\begin{equation}
-\frac{d^2\Theta}{dx^2}+\frac{1}{2}\sin2\Theta\left(\frac{d\Phi}{dx}\right)^2
-2\kappa\sin^2\Theta\frac{d\sin\Phi}{dx}%+B\sin\Theta-A\sin2\Theta
+\frac{\partial U}{\partial\Theta}=0, 
\label{eq:theta-EOM}
\end{equation}
\begin{equation}
-\frac{d}{dx}\left(\sin^2\Theta\frac{d\Phi}{dx}\right)
+2\kappa\sin^2\Theta\cos\Phi\frac{d\Theta}{dx}=0.
\label{eq:phi-EOM}
\end{equation}
These equations are written down for the case $B=0$ and $\alpha=\frac{3\pi}{2}$ in~\cite{Chovan}, where they are numerically studied, including beyond the constant $\Phi$ approximation that we make here\footnote{As a comment on notation in~\cite{Chovan} they use $\lambda$ for the DM parameter $\kappa$ and their $\gamma^{2}=2A$ is equivalent to our anisotropy parameter.}. It is worth reiterating here that the flat spirals we consider here have $\Phi(x)=$ constant, while a non-flat spiral would not have this restriction. 

As a first step, we solve Eq.~\eqref{eq:phi-EOM} for $\Phi(x)$ 
by assuming a constant value. We then find
\begin{equation}
\Phi(x)=\left(2n\pm \frac{1}{2}\right)\pi, \quad n\in\mathbb{Z}. 
\label{eq:phi-solution}
\end{equation}
Then the DM interaction term $\kappa$ disappears from the 
field equation
for $\Theta(x)$ and Eq.~\eqref{eq:theta-EOM} becomes
\begin{equation}
\frac{d^2\Theta}{dx^2}=%+B\sin\Theta-A\sin2\Theta
+\frac{\partial U}{\partial\Theta}. 
\label{eq:theta-EOM2}
\end{equation}
If we regard $x$ as ``time'', Eq.~\eqref{eq:theta-EOM2} is 
the equation of motion of a particle with unit mass moving in the 
periodic potential $-U(\Theta)$. 
Since the potential is periodic, the particle is constrained 
to move on a circle, $S^1$. This classical mechanics 
analogy is useful for classifying solutions and finding 
their properties. 
Multiplication of \eqref{eq:theta-EOM2} by $\frac{d\Theta}{dx}$ gives 
a conservation law corresponding to translational invariance 
\begin{equation}
\frac{1}{2}\left(\frac{d\Theta}{dx}\right)^{2}-U(\Theta)
%+B\cos\theta-A\cos^{2}\theta
=C_{0}, \label{eq: first integral}
\end{equation}
where the constant $C_0$ is a conserved quantity (an integral of motion) 
corresponding to the ``mechanical energy'' (sum of kinetic and ``potential energy'' $-U(\Theta)$) in classical mechanics.

\subsection{Classification of solutions }\label{sec:classification}

%We tabulate the values of the minimum $U_{\rm min}$ of the potential 
%$U(\Theta)$ and their positions $\Theta_{\rm min}$ in the angular 
%coordinate for regions I, II, and III in Table~\ref{tbl:potential_min}. 
We %also 
denote the value of the maximum of the potential as $U_{\rm max}$.

%%%%%%%%
%\begin{table}[htbp]\begin{center}\caption{\label{tbl:potential_min} 
%Values of potential at minimum $U_{\rm min}$ and their position 
%$\Theta_{\rm min}$ for regions I, II, and III. }
%\begin{tabular}{|c|c|c|c|}
%\hline
% & region I & region II & region III \\ \hline
%Parameter & $-2A\le B \le 2A$  & $A\le 0, B\ge0$ 
%&  $A\le 0, B\le0$  \\ region &   & or $0\le 2A\le B$ & or $B\le -2A\le 0$ 
%\\ \hline Ground st. & $n_3=\frac{B}{2A}$ & $n_3=1$ & $n_3=-1$ \\ \hline
%$U_{\rm min}$ & $-\frac{B^2}{4A}$ & $A-B$ & $A+B$ \\ \hline
%$\Theta_{\rm min}$ & $\pm\arccos\frac{B}{2A}+2{\mathbb Z}\pi$ 
%& $2{\mathbb Z}\pi$ & $(2{\mathbb Z}+1)\pi$ \\ \hline
%\end{tabular}\end{center}\end{table} 
%%%%%%%%%

%\begin{figure}[htbp]\begin{center}\includegraphics[width=0.6\textwidth]
%{figures/mechanical_energy_anisotropy_dominated_3}
%\caption{A plot of the potential $-U_{\rm min}(\Theta)$ for the 
%``mechanical analog" when $B<2A$. The upper solid horizontal line (red) 
%is $-U_{\rm min}=\frac{B^2}{2A}$ and the lower horizontal dashed line 
%(blue) is $-U_{\rm max}$. Motions (from right to left) with the 
%``mechanical energy" $C_0>-U_{\rm min}$ give spiral solutions. 
%Motions with $C_0=-U_{\rm min}$ connect two neighboring $\Theta_{\rm min}$ 
%and give domain wall solutions. Motions with $C_0<-U_{\rm min}$ give 
%oscillating solutions. }
%\label{fig:mech_analogy_region1}\end{center}\end{figure}

\begin{figure}[htbp]
\begin{center}\includegraphics[width=0.6\textwidth]{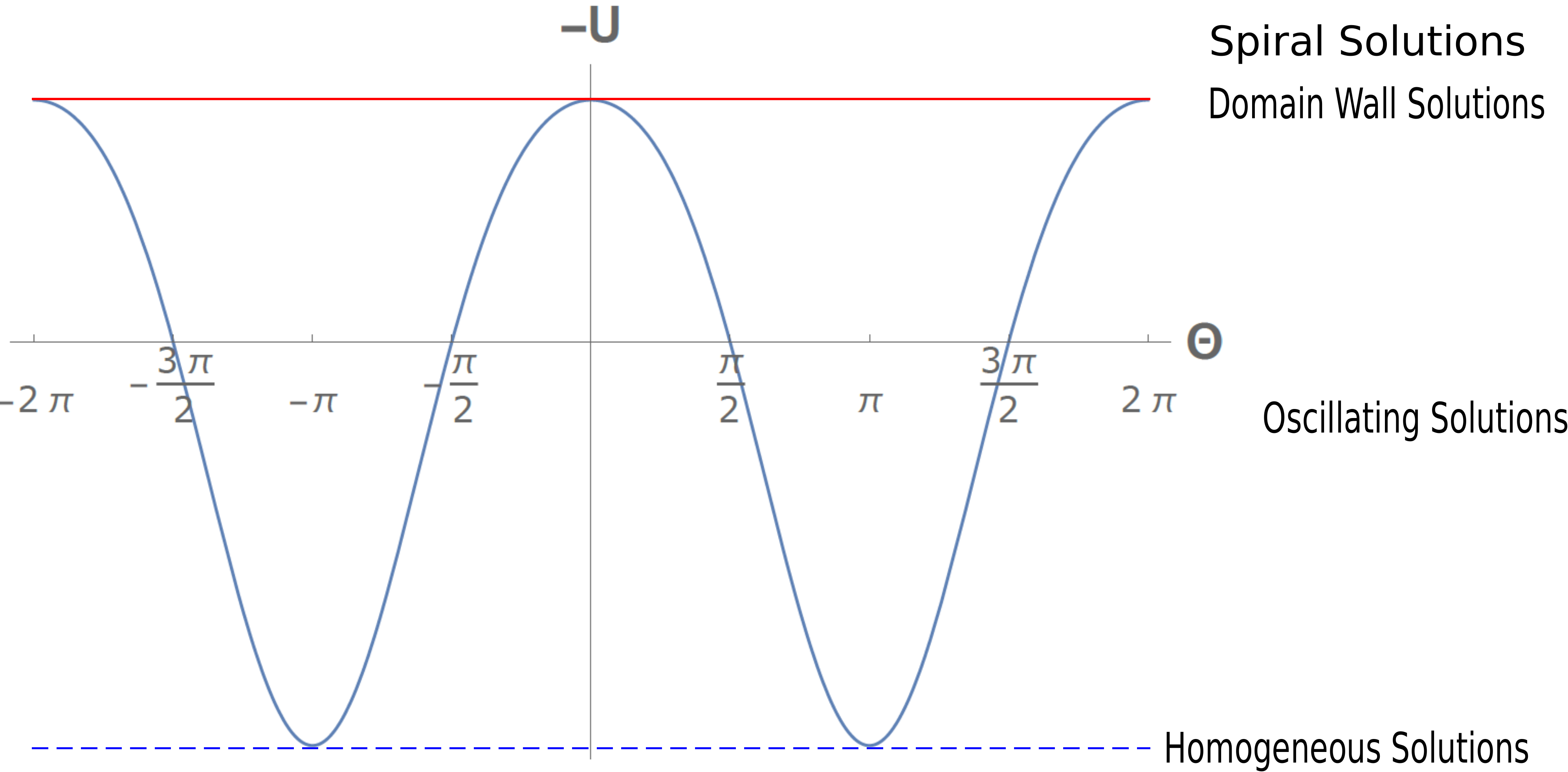}
\caption{A plot of the potential $-U_{\rm min}(\Theta)$ for 
the ``mechanical analog" when $B>2A$. The 
upper solid horizontal line (red) is $-U_{\rm min}=-A+B$ 
and the lower horizontal dashed line (blue) is $-U_{\rm max}$.
Motions (from right to left) with the ``mechanical energy" 
$C_0>-U_{\rm min}$ give spiral 
solutions. Motions with $C_0=-U_{\rm min}$ connect two neighboring 
$\Theta_{\rm min}$ and give domain wall solutions. 
Motions with $C_0<-U_{\rm min}$ give oscillating solutions. 
}
\label{fig:mech_analogy_region2}
\end{center}
\end{figure}

Since a symmetry $B\to -B, n_3\to-n_3$ allows us to obtain 
solutions for $B\le 0$ from those of $B\ge0$, we concentrate 
here on regions I and II. 
The conservation law implies that solutions exist if and only if 
$C_0\ge -U_{\rm max}$, because of the positivity $(\frac{d\Theta}{dx})^2\ge 0$. 
We depict a typical figure %s 
of $-U(\Theta)$ for the region %s I and 
II in Fig.%s.
~%\ref{fig:mech_analogy_region1} and 
\ref{fig:mech_analogy_region2}. %, respectively. 
In the captions we discuss the qualitatively different  
%we can exhaust all possible solutions of the field equation 
%\eqref{eq:theta-EOM2}, and classify qualitatively different 
types of solutions in terms of the value of $C_0$. 
%\begin{enumerate}
%\item
%$C_0 > -U_{\rm min}$ : 
%The solution is a monotonic function $\Theta(x)$. 
%We call this class of solutions spiral solutions.
%In terms of the magnetization vector, the spiral solutions are periodic.
%\item
%$C_0=-U_{\rm min}$ : 
%There are two types of solutions. 
%One of them is the homogeneous solution 
%$\Theta(x)=\Theta_{\rm min}=$constant, sitting at the top of 
%$-U(\Theta)$. 
%This is the homogeneous ground state. 
%The other is the domain wall solution connecting two neighboring 
%$\Theta_{{\rm min}}$. 
%This can be obtained as the infinite period limit of a spiral solution.
%
%\item
%$-U_{\rm max} < C_0<-U_{\rm min}$ : 
%Oscillating solutions periodically moving between two turning 
%points $\Theta_1$ and $\Theta_2$ defined by $C_0+U(\Theta)=0$. 
%%as illustrated in Fig.3. 
%For a parameter region $2A<-B<0$ in region II, there are local 
%maxima of $U(\Theta)$ at $\Theta=(2{\mathbb Z}+1)\pi$. 
%Hence there is also a bounce solution \cite{Condat1983} as an 
%infinite period limit of the oscillating solution. 
%There is also a homogeneous solution sitting at this local minimum. 
%Although this solution is stable against small fluctuations, 
%its energy has a finite positive energy gap above the global 
%minimum $U_{\rm min}$ (for $C_0<-U_{\rm min}$) and cannot give 
%a ground state. 
%\item
%$C_0=-U_{\rm max}$ : 
%Homogeneous solution $\Theta(x)=\Theta_{\rm max}$=constant, sitting 
%at the global maximum of $U(\Theta)$. 
%This is a homogeneous solution which is unstable against 
%infinitesimal fluctuations. 
%\end{enumerate}

Using the constant solution \eqref{eq:phi-solution} for $\Phi$, 
the energy density becomes 
\begin{equation}
\mathcal{E}_{\rm sol}=\kappa\frac{d\Theta}{dx}
+\frac{1}{2}\left(\frac{d\Theta}{dx}\right)^{2}+U(\Theta).
%-B\cos\Theta+A\cos^{2}\Theta 
\label{eq:energy-density-Phi-solution}
\end{equation}
The energy of the solution depends on the DM interaction term, 
even though the field equation \eqref{eq:theta-EOM2} does not. 
However, the DM term does not contribute to the 
energy for homogeneous oscillating solutions 
%($-U_{\rm max} <C_0<-U_{\rm min}$) including the bounce solutions, 
after averaging over one period.

The inequality $\frac{1}{2}\left(\frac{d\Theta}{dx}\right)^{2}
+U(\Theta)\ge U_{\rm min}$ implies that the energy density of 
the homogeneous ground state at $\Theta_{\rm min}$ is lower than 
the average energy densities of the oscillating solutions and other 
homogeneous solutions. 
We will study spiral solutions and domain wall solutions 
in order to find the ground state of a one dimensional chiral 
magnet in the following.

For the spiral and domain wall solutions with $C_0>-U_{\rm min}$, 
Eq.~\eqref{eq: first integral} implies that $\frac{d\Theta}{dx}$ 
never vanishes. Hence its sign never changes, 
and the solutions are monotonic. 
Since the energy density \eqref{eq:energy-density-Phi-solution} 
favors the negative sign for $\frac{d\Theta}{dx}$, we should choose 
the negative chirality solution with $\frac{d\Theta}{dx}<0$
in order to obtain the ground state of the chiral magnet. 
Therefore we need to solve the following first order equation 
\begin{equation}
\frac{d\Theta}{dx}=- \sqrt{2(C_0+U(\Theta))}. 
 \label{eq: theta-equation}
\end{equation}
By making use of the conservation of ``mechanical energy'' 
from Eq.~\eqref{eq: first integral}, we can rewrite the energy density 
 as 
\begin{equation}
\mathcal{E}_{\rm sol}=\kappa\frac{d\Theta}{dx}
+\left(\frac{d\Theta}{dx}\right)^{2}-C_0
=\kappa\frac{d\Theta}{dx}+2U(\Theta)+C_0. 
\label{eq:energy-density-solution}
\end{equation}

Explicit solutions are found by integrating the first order 
equation \eqref{eq: theta-equation}. 
We observe that the solutions are given in terms of 
elliptic functions. 
In the next subsection, however, we derive general properties of 
spiral solutions and will calculate the phase boundary 
between the spiral and 
homogeneous phases without needing to find explicit solutions.

\section{Properties of spiral solutions}
\label{sec:spiral_solution}

\subsection{Spiral solutions without the potential}
\label{sec:spiral_solution_no_pot}

\begin{figure}[htbp]
\begin{center}
\includegraphics[width=0.45\textwidth]{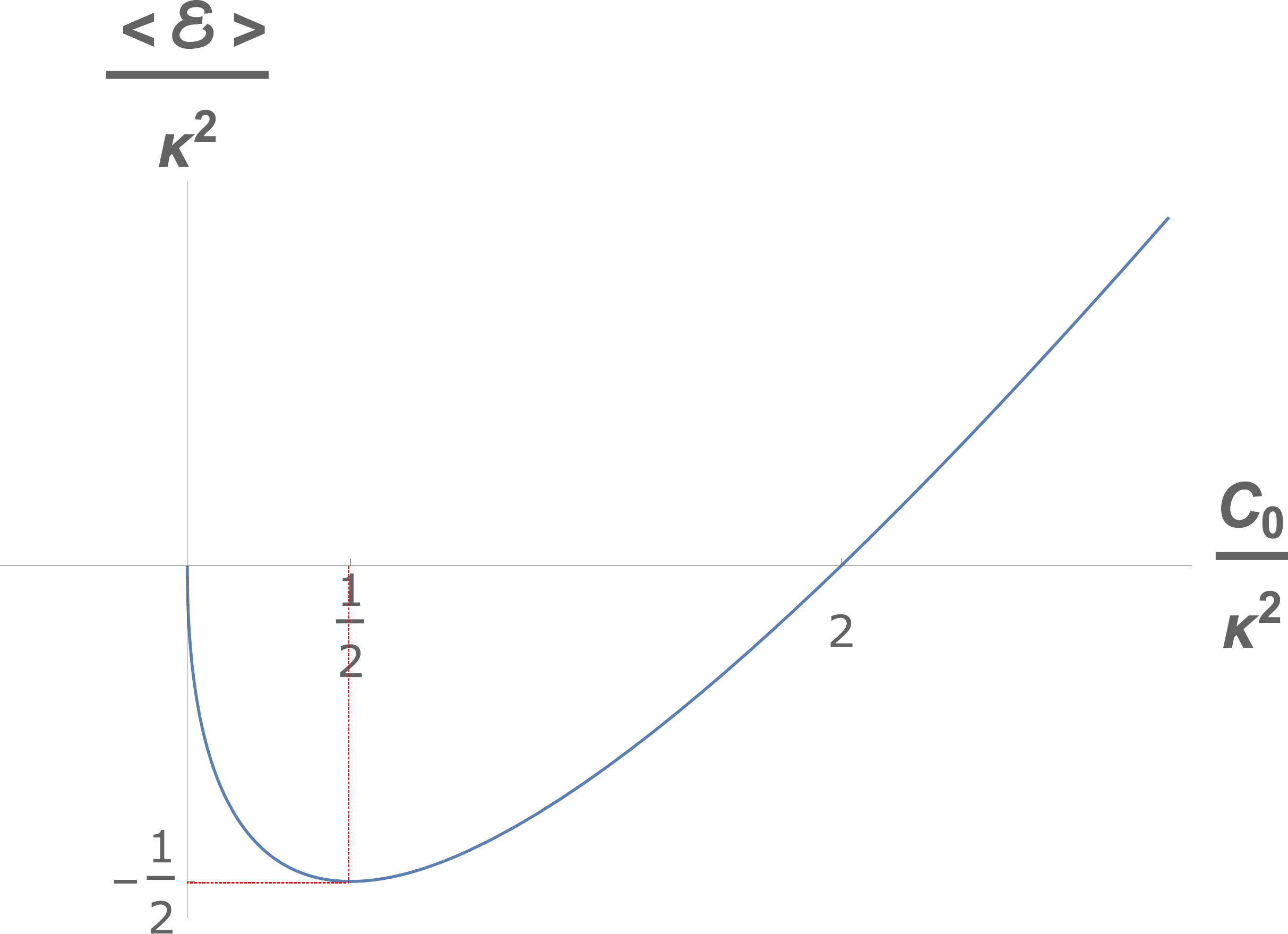}
\includegraphics[width=0.45\textwidth]{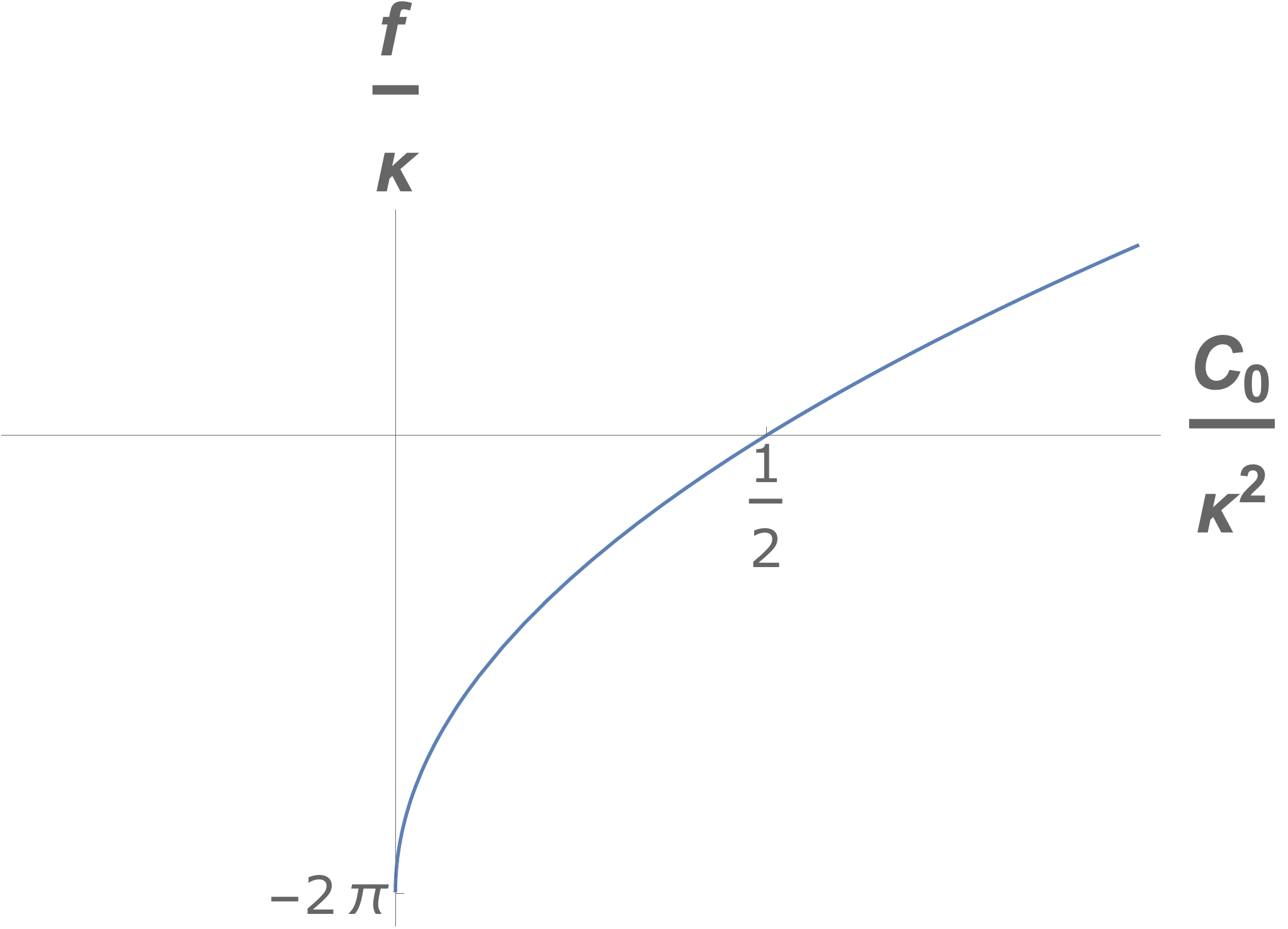}
\caption{Left: Energy density $\mathcal{E}$ of 
spiral solutions as a function of $C_{0}$ for $A=B=0$. 
The lowest energy spiral solution occurs at $C_0=C_{0, {\rm min}}=\kappa^2/2$, 
giving the energy density $\mathcal{E}_{\rm min}=-\kappa^2/2$ 
for the spiral ground state. 
Right: The excess energy $f(C_0; A=B=0)$ in one period as a function 
of $C_{0}$ for $A=B=0$. It increases monotonically and 
vanishes at $C_0=C_{0, {\rm min}}=\kappa^2/2$.} 
\label{fig:C0-E_AB=0}
\end{center}
\end{figure}

As a warm up for studying more general spirals let us first consider 
the case without a potential, $A=B=0, U(\Theta)=0$. This has been treated in the literature before \cite{BH} and we discuss it here to set up our conventions for future sections.
In this case, we need $C_0\ge 0$ and obtain $\frac{d\Theta}{dx}=-\sqrt{2C_0}$ 
from \eqref{eq: theta-equation}. The general solutions 
contain one additional parameter, $x_0$ as a Nambu-Goldstone 
(NG) mode for the spontaneously broken translational symmetry,
\begin{equation}
\Theta(x)= -\sqrt{2C_0} (x-x_0).
\label{eq:plane-wave}
\end{equation}
The magnetisation vector is
$n_3+in_2^\alpha=e^{-i\sqrt{2C_0}(x-x_0)}$ rotating along $x$ with the 
momentum $ -\sqrt{2C_0}<0$, namely a negative chirality plane wave as 
a spiral solution. 
The period $L$ of this spiral solution is
$L=2\pi/\sqrt{2C_0}$. 
Although all the configurations with different values of $C_0$ are solutions of the equations of motion, 
they have different energy densities
\begin{equation}
\mathcal{E}=-\kappa\sqrt{2C_0}+C_0. 
\label{eq:energy-density-no-pot}
\end{equation}
As illustrated in the left panel of Fig.~\ref{fig:C0-E_AB=0}, 
the energy $\mathcal{E}$ of spiral solutions as a function 
of $C_0$ has a minimum 
%$\frac{\partial\mathcal{E}}{\partial C_0}=0$ 
at $C_{0,{\rm min}}=\frac{\kappa^{2}}{2}$ 
giving the lowest energy spiral solution as 
\begin{equation}
\mathcal{E}_{\rm min}=-C_{0,{\rm min}}=-\frac{\kappa^2}{2}.  
\end{equation}

This energy density is lower than that of the homogeneous 
solution ($\mathcal{E}_{\rm homogeneous}=0$), thus
$A=B=0$ is in the spiral phase. 
For later use, we define the excess energy $f(C_0; A=B=0)$ in one 
period above the constant energy density $-C_0$ as 
\begin{equation}
f(C_0; A=B=0) =\int_0^{L} dx (\mathcal{E}+C_0)
%=\frac{2\pi}{\sqrt{2C_0}}\left(-\kappa\sqrt{2C_0}+2C_0\right)
=-2\pi\kappa+2\pi\sqrt{2C_0}. 
\label{eq:f-no-pot}
\end{equation}

As illustrated in the right panel of 
Fig.~\ref{fig:C0-E_AB=0}, $f(C_0; A=B=0)$ 
increases 
monotonically and vanishes at $C_0=C_{0, {\rm min}}=\frac{\kappa^{2}}{2}$.  

\subsection{Lowest energy flat spiral solution }
\label{sec:lowest_spiral_solution}

Although spiral solutions (including domain wall solutions) 
are monotonic functions of $\Theta$, 
they are periodic as a function of magnetization vector $\vec{n}$. 
We define the period $L$ of spiral solutions as the distance 
corresponding to translation by $2\pi$ in $\Theta$. Since 
Eq.~\eqref{eq: theta-equation} gives a one-to-one correspondence 
between $x$ and $\Theta$, once the NG mode $x_0$ is fixed 
by $\Theta(x_0)=0$ for instance. 
We can now change variable from $x$ to $\Theta$ by using 
Eq.~\eqref{eq: theta-equation}. The period $L$ of a spiral 
solution is given by 
\begin{equation}
L=\int_{0}^{L}dx=\int_0^{2\pi}d\Theta\left[-\frac{d\Theta}{dx}\right]^{-1}
=\int_{0}^{2\pi}  \frac{d\Theta}{\sqrt{2(C_0+U(\Theta))}}. 
\label{eq:period}
\end{equation}
The average energy density $\langle\mathcal{E}\rangle$ is given by 
\begin{equation}
\langle\mathcal{E}\rangle=\frac{1}{L}\int_{0}^{L}dx\,\mathcal{E}
=\frac{f(C_0;A,B)}{L}-C_0, 
\label{eq:energy-solution}
\end{equation}
where 
%\begin{widetext}
\begin{equation}
\begin{split}
f(C_0;A,B)&=\int_{0}^{L}dx\left[\kappa\frac{d\Theta}{dx}+
\left(\frac{d\Theta}{dx}\right)^2\right]\\
&=
-2\pi\kappa +\int_{0}^{2\pi} d\Theta 
\sqrt{2(C_0+U(\Theta))},
\end{split}
\label{eq:energy-solution2}
\end{equation}
%\end{widetext}
is the energy in one period due to the excess energy 
density above the constant value $-C_0$. As in Eq.~\eqref{eq:f-no-pot} we call $f(C_0;A,B)$ 
the excess energy. 
Only the first term of the excess energy comes 
from the DM interaction, this gives a negative energy contribution only when 
the solution has a non-zero winding ($-2\pi$) after one 
period of translation. 
This feature makes it a possibility for spiral solutions to have  
lower average energy density than the homogeneous ground state.

For fixed values of parameters $\kappa, A, B$, our spiral solutions 
have two parameters (moduli) $C_0$ and $x_0$. The average energy 
density depends only on $C_0$ and is 
independent of $x_0$. Therefore we need to look for the lowest energy solution among spiral 
solutions, by minimizing the average energy density as a function 
of $C_0$. We find the identities 
\begin{equation}
\frac{\partial f(C_0;A,B)}{\partial C_0}=
\int_{0}^{2\pi}  
\frac{d\Theta}{\sqrt{2(C_0+U(\Theta))}}
=L, %\right]
\label{eq:derivative-f}
\end{equation}
\begin{equation}
\frac{\partial L}{\partial C_0}=
-\int_{0}^{2\pi}  
\frac{d\Theta}{(2(C_0+U(\Theta)))^{3/2}}, %\right]
\label{eq:derivative-L}
\end{equation}
some of which are also given in \cite{Chovan}.
Thus we obtain 
\begin{equation}
\frac{\partial \langle\mathcal{E}\rangle}{\partial C_0}
=\frac{-1}{L^2}\frac{\partial L}{\partial C_0}f(C_0;A,B).
%\left[-2\pi\kappa+\int_0^{2\pi}d\Theta \sqrt{2(C_0+U(\Theta))}\right]. 
\label{eq:derivative-average-energy}
\end{equation}
Eq.~\eqref{eq:derivative-f} shows that $f(C_0;A,B)$ is monotonically 
increasing. 
Because of \eqref{eq:derivative-L}, the average energy density 
reaches a minimum when 
\begin{equation}
f(C_0;A,B)=-2\pi\kappa+\int_0^{2\pi}d\Theta\sqrt{2(C_0+U(\Theta))}=0, 
\label{eq:minimum-energy-cond}
\end{equation}
is satisfied, provided this value of $C_{0,{\rm min}}$ is in the allowed 
region of spiral solutions $C_0 \ge -U_{\rm min}$. 
This minimum energy condition \eqref{eq:minimum-energy-cond} 
gives $C_{0,{\rm min}}$ as a function of $A$ and $B$. 
Eq.~\eqref{eq:energy-solution} implies that the average energy 
density of the lowest energy spiral solution is given by this 
$C_{0,{\rm min}}$ as 
\begin{equation}
\langle\mathcal{E}\rangle_{\rm min}
=-C_{0,{\rm min}}. 
\label{eq:min-energy-spiral}
\end{equation}
Hence the lowest energy spiral solution always has 
lower energy than the homogeneous ground state with the energy 
density $U_{\min}$ 
\begin{equation}
\langle\mathcal{E}\rangle_{\rm min}
=-C_{0,{\rm min}}\le U_{\rm min}. 
\label{eq:min-energy-spiral2}
\end{equation}
We conclude that a chiral magnet is in the spiral phase, 
once the minimum energy condition is satisfied in the allowed 
region for spiral solutions $C_0> -U_{\rm min}$.

\subsection{Boundary between spiral and homogeneous phases }
\label{sec:phase_boundary}

As given in Eq.~\eqref{eq:f-no-pot}, the function $f$ 
at $A=B=0$ becomes $f(C_0; A=B=0)=-2\pi\kappa+2\pi\sqrt{2C_0}$.  
This behaves asymptotically as
$f(C_0;A=B=0) \sim 2\pi \sqrt{2C_0} > 0$ for $C_0\to \infty$ 
whereas
$%\begin{equation}
f(C_0=-U_{\rm min}=0;A=B=0)=-2\pi\kappa^2 <0$. 
%\label{eq:f_A=B=0}\end{equation}
Hence the excess energy $f(C_0;A=B=0)$ has a zero in the 
region $C_0>-U_{\rm min}$, ensuring the existence of the lowest 
energy spiral solution for $A=B=0$.

Let us now study the case of nonvanishing values of $A,B$, such that  $2A\leq B$. 
We observe that the excess energy $f(C_0;A,B)$ at 
$C_0\gg A, B, \kappa^2$ becomes asymptotically 
\begin{equation}
f(C_0;A,B) \sim 2\pi \sqrt{2C_0} > 0. 
\label{eq:asymptotic_f}
\end{equation}
Since the excess energy $f(C_0;,B)$ is monotonically increasing, 
the zero of the excess energy $f(C_0;A,B)=0$ occurs in the 
allowed region if and only if $f(C_0=-U_{\rm min};A,B)<0$.

When $C_0$ approaches the lower bound at $-U_{\rm min}$, 
we find 
\begin{equation}
L \to \infty, 
\label{eq:infinite_period}
\end{equation}
since the integrand diverges at $\Theta = \Theta_{\rm min}$, 
as $\sqrt{C_0+U(\Theta)} \to \sqrt{\frac{U''(\Theta_{\rm min})}{2}}
|\Theta-\Theta_{\rm min}|$ with $U''=\frac{d^{2}U}{d\Theta^{2}}$. 
On the other hand, the excess energy remains finite as 
$C_0\to -U_{\rm min}$. 
Therefore the average energy density \eqref{eq:energy-solution} 
becomes 
\begin{equation}
\left.\langle\mathcal{E}\rangle\right|_{C_0=-U_{\rm min}} 
=U_{\rm min}, 
\label{eq:energy_density_Umin}
\end{equation}
which is the same as that of the homogeneous solution. 
We denote the excess energy at this $C_0=-U_{\rm min}$ as 
%\begin{widetext}
\begin{align}
E_{\rm period}(A,B)&=f(C_0=-U_{\rm min};A,B)\nonumber\\
&=-2\pi\kappa+\int_0^{2\pi}d\Theta\sqrt{2(U(\Theta)-U_{\rm min})}. 
\label{eq:excess-energy-min}
\end{align}
%\end{widetext}
Summarizing the above, we find that the chiral magnet is in the 
spiral phase if\\ $E_{\rm period}(A,B)<0$, and
in the homogeneous phase if $E_{\rm period}(A,B)>0$. 
The phase boundary between spiral phase and the homogeneous phases 
is given by 
\begin{equation}
0=E_{\rm period}(A,B)
=-2\pi\kappa+\int_0^{2\pi}d\Theta\sqrt{2(U(\Theta)-U_{\rm min})}. 
\label{eq:phase-boundary-line}
\end{equation}
Since the period is infinite, the spiral solution in this limit 
becomes a domain wall solution connecting two adjacent values 
of $\Theta_{\min}$. We will describe more explicitly the 
domain wall solutions as a limit of spiral solutions in a later 
section. 
Since $-U_{\rm min}$ is the energy density of the homogeneous solution, 
the excess energy above this energy density integrated over one period 
$E_{\rm period}(A,B)$ is precisely the domain wall energy as a 
finite energy soliton solution on the homogeneous background. 
Our results show that the domain wall solution as a single 
soliton above the homogeneous background has positive energy 
and is an ordinary excitation in the homogeneous phase. 
It becomes zero energy at the boundary between the homogeneous and 
spiral phases. 
It has a negative energy in the spiral phase region, and 
signals the instability of the homogeneous background solution 
and its decay to the lowest energy spiral solution as the true 
ground state. 
One can intuitively visualize the deformation of the configuration 
as a condensation of these negative energy solitons settling down 
to the spiral ground state.

\subsection{Explicit formula for phase boundaries}\label{sec:phaseboundaries}

%Since the values of $U_{\rm min}$ differ in the two 
%regions corresponding to the two different homogeneous 
%ground states, we need to distinguish these
%two parameter regions in order to determine the phase 
%boundary between the homogeneous phase and the
%spiral phase in 
%Eq.~\eqref{eq:phase-boundary-line}. 

\begin{figure}[htbp]
\begin{center}
\includegraphics[width=0.5\textwidth]{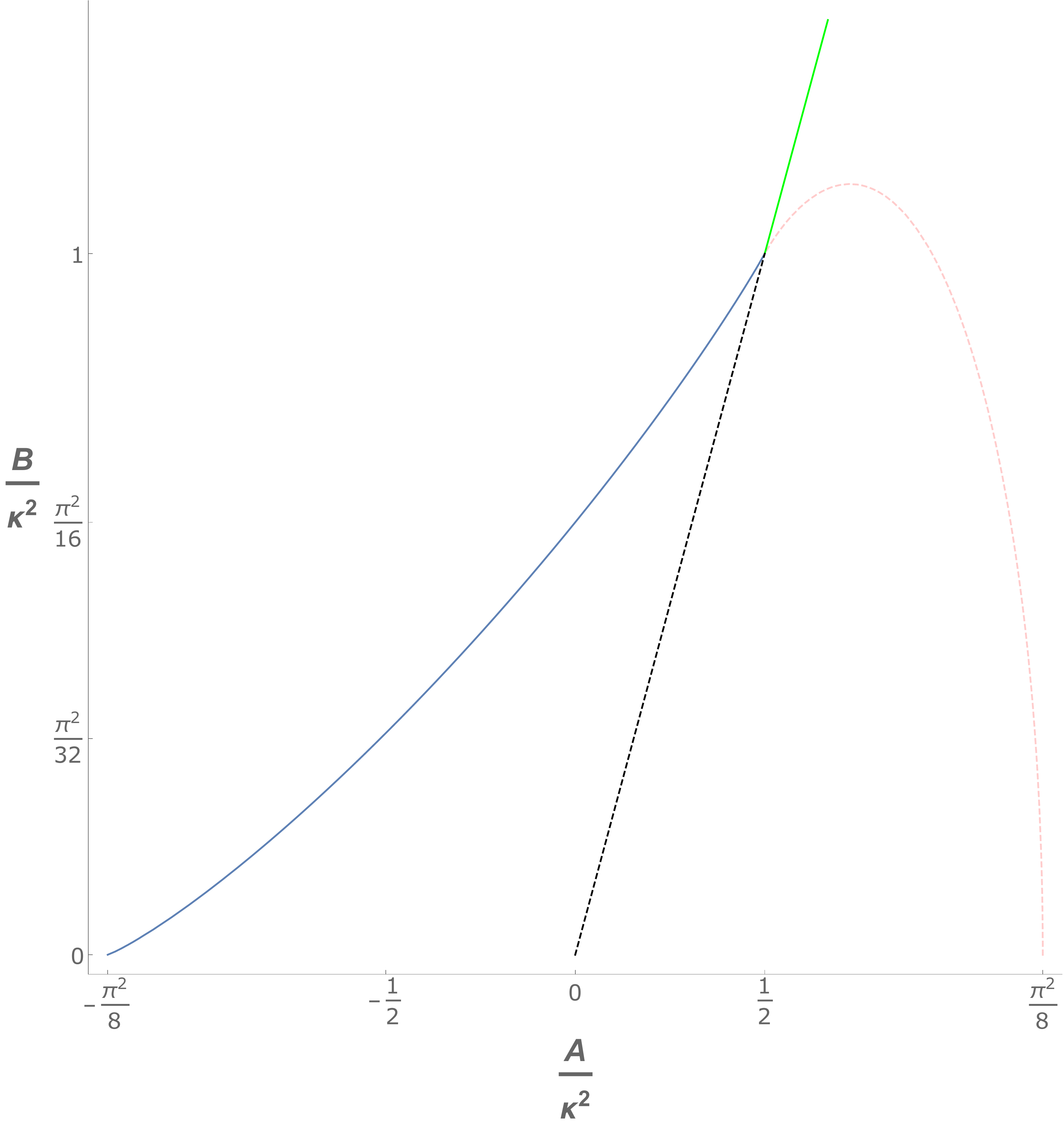}
\caption{The phase diagram in the 
$\frac{A}{\kappa^{2}}, \frac{B}{\kappa^{2}}$ plane. 
The spiral phase is below the blue curve. 
The(positively) polarized ferromagnetic phase is above 
the blue curve. The two ferromagnetic phases are separated by 
the green line $B=2A$. 
Below the tricritical point $B=2A=\kappa^2$, the dashed line $B=2A$
does not correspond to a phase boundary. The dashed red line shows where flat spirals in the anisotropy dominated case have zero energy. This is not a phase boundary as the flat spiral is not the ground state. 
}
\label{fig:Phase boundaries}
\end{center}
\end{figure}

%\begin{enumerate}
%\item
%region I : $0\le B \le 2A$ 
%
%For the homogeneous phase, this is the region of the
%canted ferromagnetic phase $n_3=\frac{B}{2A}$ corresponding to 
%$\Theta_{\rm min}=\pm\arccos(\frac{B}{2a})+2{\mathbb Z}\pi$ with the 
%minimum value $U_{\rm min}=-\frac{B^{2}}{4A}$ of the potential. 
%We find the boundary in Eq.~\eqref{eq:phase-boundary-line} between 
%the canted ferromagnetic phase and spiral phase to be
%%\begin{widetext}
%\begin{align}
%2\pi\kappa&=\int_0^{2\pi}d\Theta
%\sqrt{2\left(-B\cos\Theta+A\cos^2\Theta +\frac{B^2}{4A}\right)}
%\nonumber \\
%&=4\sqrt{2A-\frac{B^2}{2A}}
%+2\sqrt{2A}\frac{B}{A}\left(\frac{\pi}{2}-\arccos\frac{B}{2A}\right). 
%\label{eq:phase-boundary-case1}
%\end{align}
%%\end{widetext}
%This boundary curve is depicted as the (red) curve to 
%the right of the $B=2A$ line in Fig.~\ref{fig:Phase boundaries}.

%\item
When $2A\le B$, the homogeneous phase in this region is the
polarized ferromagnetic 
phase $n_3=+1$. 
The minimum of the potential is $U_{\rm min}=A-B$ which occurs at 
$\Theta_{\rm min}=2{\mathbb Z}\pi$. 
We find that the boundary in Eq.~\eqref{eq:phase-boundary-line} 
between 
the polarized ferromagnetic phase and the spiral phase is given by
\begin{align}
2\pi\kappa
&=
\int_0^{2\pi}d\Theta
\sqrt{2(-B\cos\Theta+A\cos^2\Theta -A+B)}
\nonumber \\
&=
4\sqrt{B-2A}
+\frac{4B}{\sqrt{2A}}\arcsin\sqrt{\frac{2A}{B}}. 
\label{eq:phase-boundary-case2}
\end{align}
This boundary is depicted as the (blue) curve to the left of 
$B=2A$ line in Fig.~\ref{fig:Phase boundaries}.
This expression is manifestly real for $0\le A$. 
This formula is also valid for $A\le 0$ by an analytic continuation 
of $A>0\to A<0$. 
We find a convenient (manifestly real) expression in the region $A\le 0$ 
to be
\begin{align}
2\pi\kappa
&=
4\sqrt{B-2A}
+\frac{4B}{\sqrt{-2A}}\arcsinh\sqrt{\frac{-2A}{B}}
\label{eq:phase-boundary-case2-2}
 \\
&=
4\sqrt{B-2A}
+\frac{2B}{\sqrt{-2A}}\log\left(\frac{\sqrt{B-2A}+\sqrt{-2A}}
{\sqrt{B-2A}-\sqrt{-2A}}\right), 
\nonumber
\end{align}
where the last expression agrees with the one derived in the 
case of $A\le 0$ previously \cite{BH}, apart from the difference 
in conventions. 

This boundary curve is also depicted as the (blue) curve 
to the left of the 
$B=2A$ line in Fig.~\ref{fig:Phase boundaries}.

%\end{enumerate}
For completeness, in order to see the region of negative $B$, 
we note that the phase diagram is symmetric under $B \to -B$. 

Although partial results for particular parameter values or regions 
have been obtained for the phase boundary between homogeneous 
ferromagnetic phases and spiral phase, our results give the 
complete boundary curve explicitly for the first time.

%In sect.~\ref{sec:exact_spiral_solution}, we will illustrate our observations in detail through explicitly constructing exact spiral and domain wall solutions.

\subsection{Order of phase transition between spiral and homogeneous phases}

In this subsection, we will demonstrate that the phase transition 
between the spiral and homogeneous phases is of second order.

The phase transition is of the $n$-th order if the $l$-th 
order derivatives of the free energy at the phase boundary 
are continuous for $l<n$ and discontinuous for $l=n$. 
At zero temperature, the free energy density is given by the 
average energy density of the ground state. 
In the homogeneous phase, the energy density is given by the 
minimum  $U_{\rm min}$ of the potential. 
%as given in Table~\ref{tbl:potential_min}. 
The average  energy density of spiral phase 
$\langle\mathcal{E}\rangle_{\rm spiral}$ is given by that 
of the lowest energy spiral solution in Eq.~\eqref{eq:min-energy-spiral} 
\begin{equation}
\langle\mathcal{E}\rangle_{\rm spiral}
=-C_{0,{\rm min}}(A, B), 
\label{eq:spiral_energy}
\end{equation}
where the mechanical energy $C_{0,{\rm min}}$ of the lowest 
energy spiral solution is determined as a function of the potential 
parameters $A, B$ by the minimum energy condition in 
Eq.~\eqref{eq:minimum-energy-cond} as 
%\begin{widetext}
\begin{align}
2\pi\kappa&=\int_0^{2\pi}d\Theta\sqrt{2(U(\Theta)+C_{0,{\rm min}})}\nonumber \\
&=\int_0^{2\pi}d\Theta\sqrt{2(-B\cos\Theta+A\cos^2\Theta+C_{0,{\rm min}})}
. 
\label{eq:boundary-line}
\end{align}
%\end{widetext}
As noted in Eq.~\eqref{eq:energy_density_Umin}, the energy 
density of the spiral phase tends continuously to that 
of the homogeneous phase at the phase boundary
\begin{equation}
\langle\mathcal{E}\rangle_{\rm spiral}\to 
U_{\rm min}(A, B)=\langle\mathcal{E}\rangle_{\rm homogeneous} . 
\end{equation}
Since the energy density along the boundary is common for homogeneous 
and spiral phases, all of their derivatives tangential to the 
phase boundary are continuous between the
homogeneous and spiral phases. 
Therefore, we need to consider only the derivative perpendicular 
to the boundary.

Let us parametrize the difference between the energy density of the spiral and 
homogeneous phases in terms of the deviation $\Delta C\ge 0$ defined as 
\begin{equation}
C_{0,{\rm min}}(A, B)=-U_{\rm min}(A, B)+\Delta C(A, B), 
\label{eq:DeltaC}
\end{equation}
\begin{equation}
\langle\mathcal{E}\rangle_{\rm spiral}
- 
\langle\mathcal{E}\rangle_{\rm homogeneous}
 =-\Delta C(A, B). 
\label{eq:diff_spiral_homo}
\end{equation}
The limiting procedure of the parameters $A, B$ approaching the 
phase boundary is equivalent to $\Delta C(A,B)\to 0$. 
We can obtain the first derivative of $\Delta C$ 
in terms of $A, B$ by differentiating the 
minimum energy condition in Eq.~\eqref{eq:boundary-line} 
\begin{equation}
0=\int_0^{2\pi}d\Theta
\frac{-\frac{\partial U_{\rm min}}{\partial A}
+\frac{\partial\Delta C}{\partial A}
+\cos^2\Theta}{\sqrt{2(U(\Theta)-U_{\rm min}+\Delta C)}}, 
\label{eq:1st-derivative-A}
\end{equation}
\begin{equation}
0=\int_0^{2\pi}d\Theta
\frac{-\frac{\partial U_{\rm min}}{\partial B}
+\frac{\partial\Delta C}{\partial B}
-\cos\Theta}{\sqrt{2(U(\Theta)-U_{\rm min}+\Delta C)}}. 
\label{eq:1st-derivative-B}
\end{equation}
We can rewrite these relations as
\begin{equation}
-\frac{\partial\Delta C}{\partial A}
=\frac{L_0^{01}}{L},  \quad 
-\frac{\partial\Delta C}{\partial B}
=\frac{L_0^{10}}{L}, 
\label{eq:1stBder_Ble2A}
\end{equation}
where the period $L$ and weighted integrals $L_n^{kl}$ are defined as 
\begin{equation}
L=L_0^{00}=\int_0^{2\pi}
\frac{d\Theta}{\sqrt{2(U(\Theta)-U_{\rm min}+\Delta C)}} ,
\label{eq:period-integral-rep}
\end{equation}
and
\begin{equation}
L_n^{kl}=\int_0^{2\pi}d\Theta
\frac{\left(-\frac{\partial U_{\rm min}}{\partial B}-\cos\Theta\right)^k
\left(\cos^2\Theta-\frac{\partial U_{\rm min}}{\partial A}\right)^l}
{[2(U(\Theta)-U_{\rm min}+\Delta C)]^{n+1/2}}. 
\label{eq:weighted-integral}
\end{equation}
As is derived in Appendix \ref{sec:der_spiral}, 
$L\to \infty$ in the limit of $\Delta C\to 0$, whereas 
$L_0^{10}, L_0^{01}$ are finite. 
Therefore, we obtain at the phase boundary 
\begin{equation}
\frac{\partial \Delta C
%\langle\mathcal{E}\rangle_{\rm spiral}
}{\partial A}\to 0, \quad 
\frac{\partial \Delta C
%\langle\mathcal{E}\rangle_{\rm spiral}
}{\partial B}\to 0, 
\label{eq:1st_der_continuous}
\end{equation}
implying that the first derivative of energy density 
is continuous at the phase boundary. 

Similarly we can obtain the second derivative of $\Delta C$ by 
differentiating \eqref{eq:1st-derivative-B} and 
\eqref{eq:1st-derivative-A} in terms of $A, B$ again. 
We find an exact expression for 
$\frac{\partial^2\Delta C}{\partial A_i \partial A_j}$ with $A_1=A, A_2=B$ 
using the integral in Eq.~\eqref{eq:weighted-integral} . 
As described in Appendix \ref{sec:der_spiral}, we find that the 
second derivative diverges at the boundary 
\begin{equation}
-\left(
\begin{array}{cc}
\frac{\partial^2 \Delta C}{\partial A^2} 
& \frac{\partial^2 \Delta C}{\partial A \partial B} \\
\frac{\partial^2 \Delta C}{\partial B \partial A} 
& \frac{\partial^2 \Delta C}{\partial B^2}
\end{array}
\right)
\sim 
- \frac{L_1^{00}}{L^3}
\left[
\left(
\begin{array}{c}
L_0^{01} \\
L_0^{10} 
\end{array}
\right)
\left(
\begin{array}{cc}
L_0^{01} & L_0^{10} 
\end{array}
\right)
\right]_{\rm boundary}
, 
\label{eq:C-der-bound}
\end{equation}
with the asymptotic behavior as $\Delta C\to 0$ 
\begin{equation}
\frac{L_1^{00}}{L^3}\sim \frac{1}{\Delta C\left(\log\frac{1}{\Delta C}\right)^{3}}
\to +\infty. 
\end{equation}
Therefore the second derivative of the difference of energy density 
$\langle\mathcal{E}\rangle_{\rm spiral}- 
\langle\mathcal{E}\rangle_{\rm homogeneous}=-\Delta C$ 
is not continuous. 
Hence we conclude that the phase transition between homogeneous 
phases and the spiral phase is of second order.

The rank of the coefficient matrix is unity. 
This corresponds to the fact that the derivative is nonzero only along 
the normal to the phase boundary. 
The boundary is defined by Eq.~\eqref{eq:phase-boundary-line} 
corresponding to $\Delta C=0$. 
By varying this condition we find that along the boundary 
\begin{equation}
0=\frac{dA}{dB}\left[L_1^{01}\right]_{\rm boundary}
+\left[L_0^{10}\right]_{\rm boundary}. 
\label{eq:derivative_boundary}
\end{equation}
The relation shows that the second derivative vanishes 
along the tangential direction to the boundary.

%%%%
\subsection{Exact spiral solutions}
\label{sec:exact_spiral_solution}

As we have seen in the previous sections, there is a spiral phase 
in the regions below the boundary line defined in 
Eq.~\eqref{eq:phase-boundary-line}. 
In order to illustrate our results more explicitly, we here present 
exact spiral solutions for the simple typical cases. 
The solutions are given in terms of elliptic functions and the technical details are relegated to Appendix~\ref{app:positiveA_spiral}. Here we just present the period and energy of the solutions. Throughout this section $K(k)$ and $E(k)$ are the complete elliptic integral of 
the first and the second kind defined as 
\begin{align}
K(k)&=\int_{0}^{\frac{\pi}{2}}\frac{d\varphi}{\sqrt{1-k^{2}\sin^{2}\varphi}},\\
E(k)&=\int_{0}^{\frac{\pi}{2}}d\varphi\sqrt{1-k^{2}\sin^{2}\varphi},
\end{align}
respectively. 

\subsubsection{Spirals without external magnetic fields (\texorpdfstring{$B=0$}{B=0})}
%Some of the results of this section were already found in \cite{Chovan} where the spiral phase was studied by a mixture of  numerical and analytic approaches. In \cite{Chovan} the authors 
%found the energy density and the phase transition point for the spiral solution. However, the explicit formula for the boundary is not given. There is also numerical evidence in \cite{Chovan} that near the phase boundary there exists a ``non-flat" spiral state, with $\Phi(x)$ not a constant. This non-flat spiral appears to have lower energy than the standard, ``flat" spiral state that we study here. It does not appear to be possible to study the ``non-flat" spiral analytically, thus what we present here is the phase boundary for the ``flat" spiral.

In the case of nonzero anisotropy ($A\not=0$) without 
the Zeeman term ($B=0$), the first order equation for $\Theta(x)$ 
in Eq.~\eqref{eq: theta-equation} becomes 
\begin{equation}
\frac{d\Theta}{dx}=- \sqrt{2(C_0+A\cos^2\Theta)}. 
 \label{eq: theta-equation_B=0+}
\end{equation}

%\begin{enumerate}
%\paragraph{$A>0$ case:}
%\label{sec:positiveA_spiral}
%
%The solutions are elliptic functions with the elliptic modulus $k$ 
%and the period $L$ 
%\begin{equation}
%k=\sqrt{\frac{A}{C_0+A}}, \quad 
%\label{eq:elliptic_k_B=0+}
%L=\frac{4}{\sqrt{2(C_0+A)}} K(k) ,
%% \label{eq: period_B=0+}
%\end{equation}
%The average energy density of the spiral solution 
%$\langle \mathcal{E}\rangle =\frac{f(C_0;A>0,B=0)}{L}-C_0$ 
%is minimized by the condition of vanishing excess energy 
%\begin{equation}
%f(C_0;A>0,B=0)=-2\pi\kappa +4\sqrt{2(C_0+A)}E(k)=0, 
%\label{eq:min_cond_B=0+}
%\end{equation}
%which determines $C_0(A)$ as a function of $A$ through 
%\begin{equation}
%A=\frac{\pi^{2}\kappa^{2}}{8}\left(\frac{k}{E(k)}\right)^2, 
%\label{eq:A-k-relation_B=0+}
%\end{equation}
%and \eqref{eq:elliptic_k_B=0+}. 
%Then the minimum energy density is given by 
%\begin{equation}
%\langle \mathcal{E}\rangle
% =-C_0(A). \label{eq:lowest_energy_spiral_B=0+}
%\end{equation}
%This energy density $-C_0(A)\le 0$ is lower than the energy density of the 
%canted ferromagnetic state $\mathcal{E}_{\rm canted}=U_{\rm min}=0$. 
%As $C_0$ approaches the lowest allowed value $-U_{\rm min}=0$, 
%the period $L$ becomes infinite, 
%and the lowest energy spiral solution has the same average energy 
%density as the canted ferromagnetic state. 
%This critical point $A=\frac{\pi^{2}}{8}\kappa^{2}$ occurs 
%at the $C_0\to 0$ limit of the minimum energy condition 
%\eqref{eq:min_cond_B=0+}. 

\begin{figure}[htbp]
\begin{center}
\includegraphics[width=0.45\textwidth]
{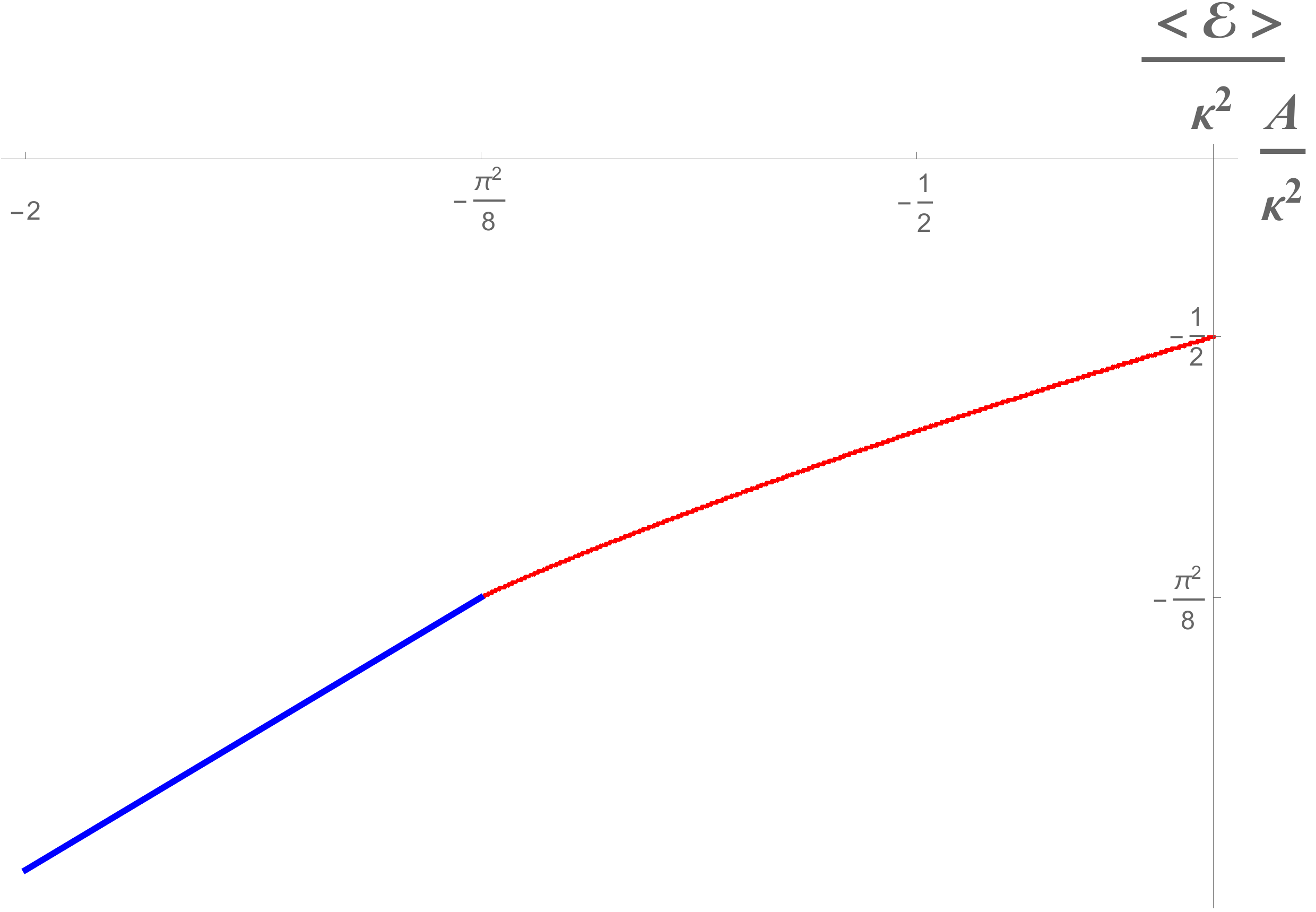}
\includegraphics[width=0.45\textwidth]
{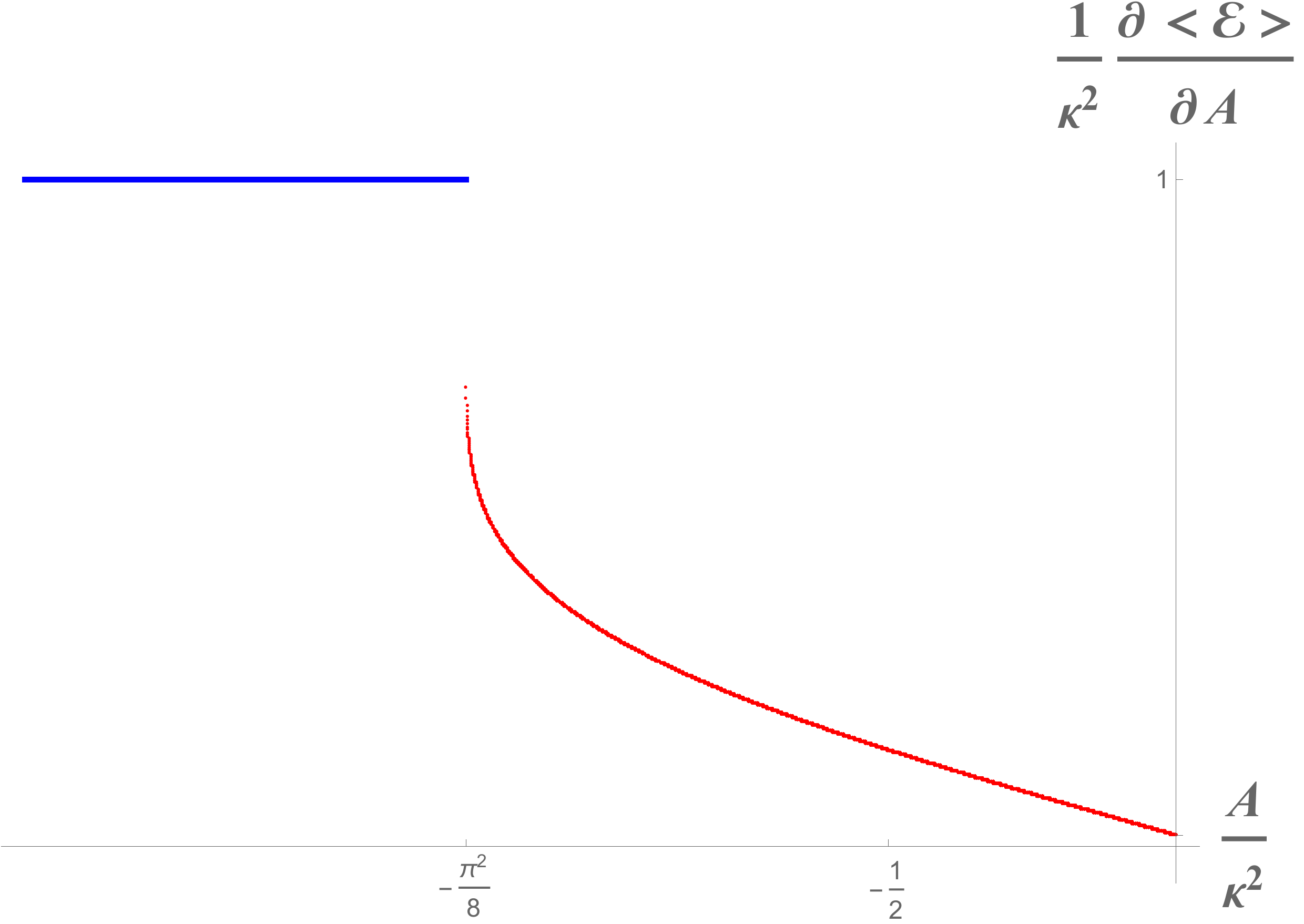}
\caption{Case of $B=0$. Red solid curves and blue solid curves correspond to
spiral phase and homogeneous ferromagnetic phase, respectively. 
Left: Average energy density 
$\langle \mathcal{E}\rangle_{min}=-C_{0}(A,0)$ as a function 
of $\frac{A}{\kappa^{2}}$, determined by Eqs.~\eqref{eq:elliptic_k_B=0-}, 
\eqref{eq:A-k-relation_B=0-}, and 
\eqref{eq:lowest_energy_spiral_B=0-} for $A<0$. 
Right: Derivative of average energy density as a function of 
$\frac{A}{\kappa^{2}}$. 
The second derivative diverges at the phase boundary 
$A=- \frac{(\pi\kappa)^{2}}{8}$ showing the second order phase transition. 
}
\label{fig:spiral_energy_density_B=0}
\end{center}
\end{figure}

\paragraph{$A<0$ case:}

The elliptic modulus $k$ and the period $L$ in this case are given by 
\begin{equation}
k=\sqrt{\frac{-A}{C_0}}, \quad 
\label{eq:elliptic_k_B=0-}
L=\frac{4}{\sqrt{2C_0}} K(k) .
% \label{eq: period_B=0-}
\end{equation}
The average energy density of the spiral solution 
$\langle \mathcal{E}\rangle =\frac{f(C_0;A>0,B=0)}{L}-C_0$ 
is minimized by the condition of vanishing excess energy 
\begin{equation}
f(C_0;A<0,B=0)=-2\pi\kappa 
+4\sqrt{2C_0}E(k)=0 ,
 \label{eq: excess_energy_B=0-}
\end{equation}
which determines $C_0(A)$ as a function of $A$ through 
\begin{equation}
A=-\frac{\pi^{2}\kappa^{2}}{8}\left(\frac{k}{E(k)}\right)^2. 
\label{eq:A-k-relation_B=0-}
\end{equation}
and \eqref{eq:elliptic_k_B=0-}. 
The minimum energy density 
\begin{equation}
\langle \mathcal{E}\rangle%_{\rm min}
 =-C_0(A). 
\label{eq:lowest_energy_spiral_B=0-}
\end{equation}
The critical point occurs at $A=-\frac{\pi^2}{8}$. 
At this point, the energy density of the lowest energy spiral solution 
becomes equal to that of the ferromagnetic state. 
%\end{enumerate}

The average energy density of the lowest energy spiral solutions 
and its derivatives are plotted in Fig.~\ref{fig:spiral_energy_density_B=0}
exhibiting the second order phase transitions at 
$A=\pm \frac{\pi^{2}\kappa^{2}}{8}$.

\subsubsection{Spiral solutions without anisotropy (\texorpdfstring{$A=0$}{A=0})}
The solution for $A=0$ has been studied previously~\cite{KO2015}. 
In this case the first order equation for $\Theta(x)$ in Eq.~\eqref{eq: theta-equation} 
becomes 
\begin{equation}
\frac{d\Theta}{dx}=- \sqrt{2(C_0-B\cos\Theta)}. 
 \label{eq: theta-equation_A=0}
\end{equation} 

The solutions are elliptic functions with the elliptic modulus $k$ 
and the period $L$ 
\begin{equation}
k=\sqrt{\frac{2B}{C_0+B}}, \quad 
\label{eq:elliptic_k_A=0}
L=2\sqrt{\frac{2}{C_0+B}} K(k) .
% \label{eq: period_A=0}
\end{equation}
The average energy density of the spiral solution 
$\langle \mathcal{E}\rangle =\frac{f(C_0;A>0,B=0)}{L}-C_0$ 
is minimized by the condition of vanishing excess energy 
\begin{equation}
f(C_0;A=0,B)=-2\pi\kappa 
+4\sqrt{2(C_0+B)}E(k) ,
 \label{eq: excess_energy_A=0}
\end{equation}
which determines $C_0(B)$ as a function of $B$ through 
\begin{equation}
B=\frac{\pi^{2}\kappa^{2}}{16}\left(\frac{k}{E(k)}\right)^2, 
\label{eq:B-k-relation_A=0}
\end{equation}
and \eqref{eq:elliptic_k_A=0}. 
The minimum energy density is given by 
\begin{equation}
\langle \mathcal{E}\rangle =-C_0(B), 
\label{eq:lowest_energy_spiral_A=0}
\end{equation}
where $C_0(B)$ is defined by \eqref{eq: excess_energy_A=0}. 
This energy density $-C_0(B)\le 0$ is lower than the energy density of 
the positively polarized ferromagnetic state with the 
energy density $-B$. 
As $C_0$ approaches the lowest allowed value $-U_{\rm min}=B$, 
the period $L$ becomes infinite, and the 
lowest energy spiral solution has the same average energy 
density as the ferromagnetic state. 
This critical point $B=\frac{\pi^{2}}{16}\kappa^{2}$ occurs 
at the $C_0\to 0$ limit of the minimum energy condition 
\eqref{eq: excess_energy_A=0}.

\begin{figure}[htbp]
\begin{center}
\includegraphics[width=0.45\textwidth]
{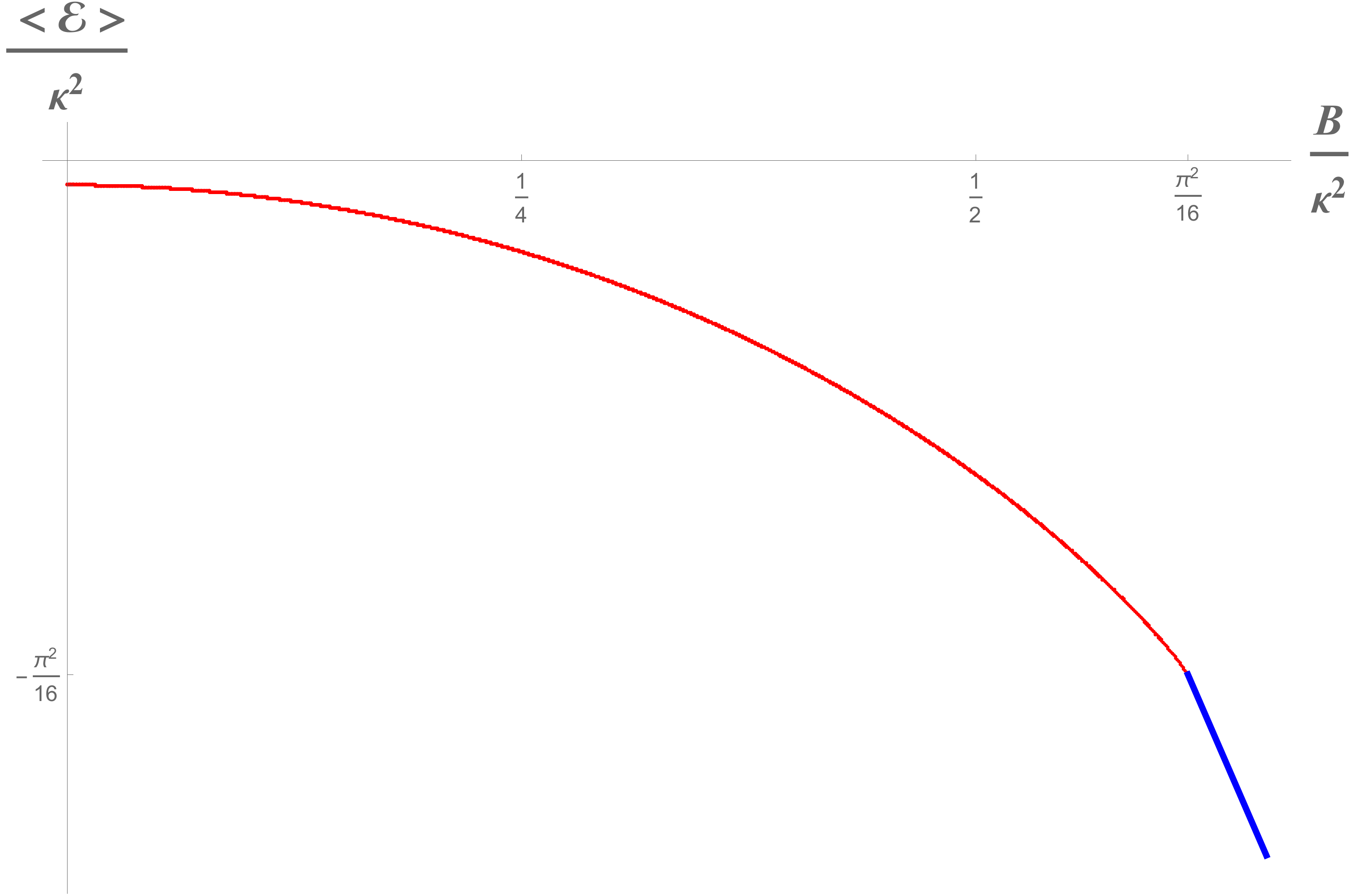}
\includegraphics[width=0.45\textwidth]
{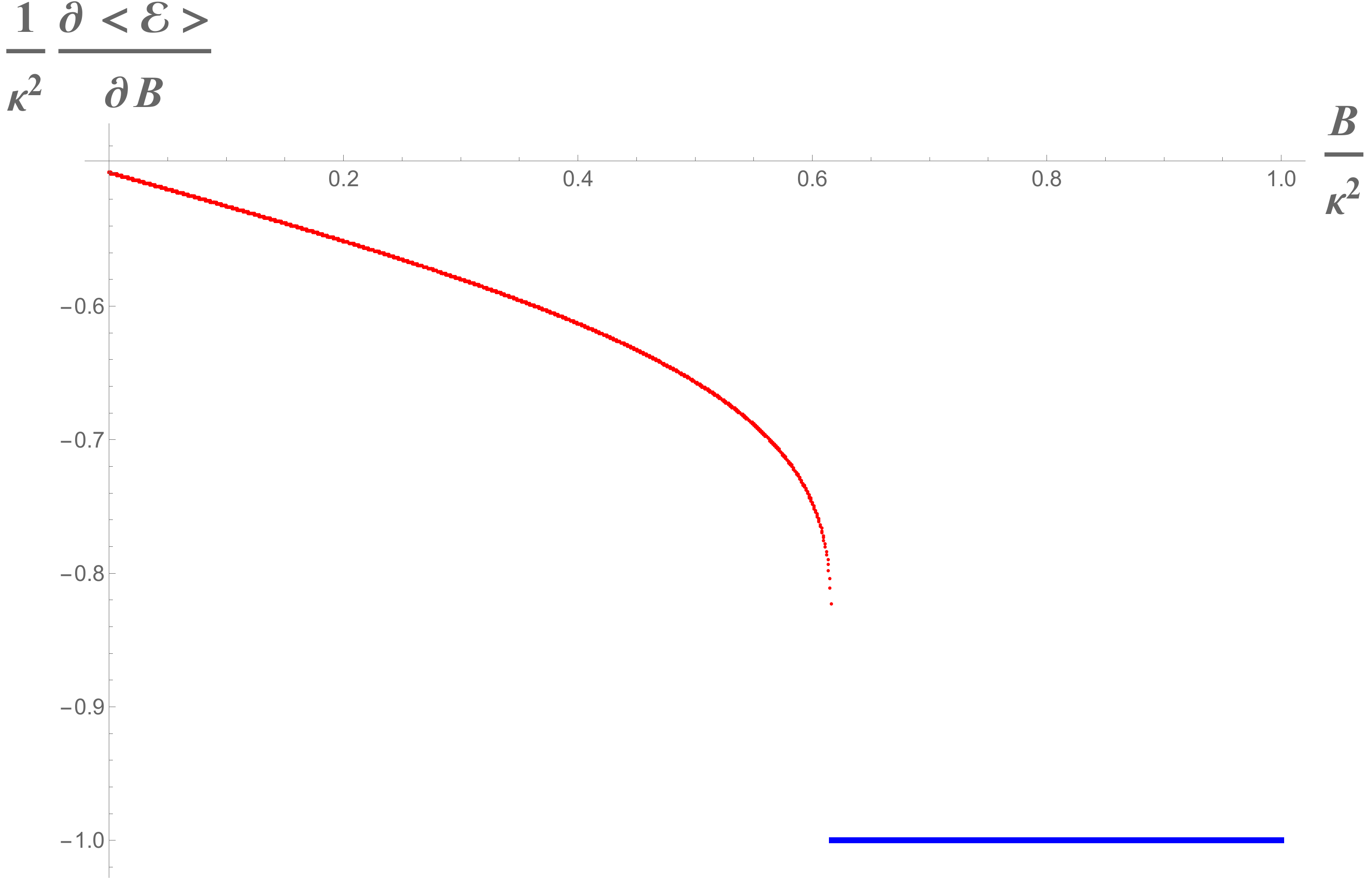}
\caption{Case of $A=0$. Red solid curves and blue solid curves correspond to
spiral phase and homogeneous ferromagnetic phase, respectively. 
Left: Average energy density 
 $\langle \mathcal{E}\rangle_{min}=-C_{0}(A=0,B)$ as a function 
of $\frac{B}{\kappa^{2}}$, determined by Eqs.~\eqref{eq:elliptic_k_A=0}, 
\eqref{eq:B-k-relation_A=0}, and \eqref{eq:lowest_energy_spiral_A=0}. 
Right: Derivative of average energy density as a function of 
$\frac{B}{\kappa^{2}}$. 
The second derivative diverges at the phase boundary 
$B= \frac{(\pi\kappa)^{2}}{16}$ showing the second order phase transition. 
}

\label{fig:min average energy density}
\end{center}
\end{figure}

As an example we draw the energy density, its first and second derivatives 
as a function of $A$ in the case of $B=0$ in
Fig.~\ref{fig:spiral_energy_density_B=0}.
Similar plots of the energy density, and its first and second 
derivatives are given as a function of $B$ for the case of 
$A=0$ in Fig.~\ref{fig:min average energy density}.

\subsection{Domain walls as infinite period limit of spirals}

For generic values of parameters $A, B$, one can still integrate 
\eqref{eq: theta-equation}, since it is an elliptic integral. 
We can express the one-to-one mapping between $x$ and $\Theta$ 
of spiral solutions in terms of Jacobi elliptic functions, 
although the mapping is similar to, but algebraically more involved 
than the explicit spiral solutions in simple cases: 
Eq.~\eqref{eq: solution_ellipticFunc_A=0} in the case of $A=0$, 
%Eq.~\eqref{eq: solution_ellipticFunc_B=0+} in the case of $B=0, A>0$, 
and Eq.~\eqref{eq: solution_ellipticFunc_B=0-} in the case of $B=0, A<0$. 
One should note that the functional form of the profile function
$\Theta(x)$ of the 
domain wall solutions or spiral solutions are determined solely 
by the parameters of the potential $A,B$ and $C_0$, but are 
independent of the DM interaction parameter $\kappa$. 
The DM interaction parameter comes in only when we evaluate 
the average energy density of the spiral solutions or domain wall solutions. 
Therefore the lowest energy spiral solutions and the critical point 
explicitly depends on the DM parameter $\kappa$. 
Similarly the domain wall solutions can exist irrespective of the value 
of $\kappa$, but they become zero energy solitons at the 
phase boundary between spiral and homogeneous ferromagnetic 
phases.

Let us examine the infinite period limit of the lowest energy 
spiral solutions. 
As we approach the phase boundary, the period of spiral solutions 
tends to infinity, and the solutions become domain wall solutions, 
as we argued. 
At the boundary we find the average energy density 
becomes identical to that of the homogeneous solution. 
Moreover, the total energy of the domain wall vanishes at the boundary. 
We will explicitly see these properties by studying domain wall 
solutions in all parameter regions across the phase boundary 
in the following section. 
The domain wall solutions $\Theta(x)$ can be expressed in terms 
of elementary functions. 
This corresponds to the fact that the elliptic function describing 
a spiral solution becomes an elementary function in the limit 
of infinite period.
Let us take the limit of infinite period for illustrative examples 
of spiral solutions in the case of $A=0$ and $B=0$.

\begin{figure}[htbp]
\begin{center}
\includegraphics[width=0.45\textwidth]{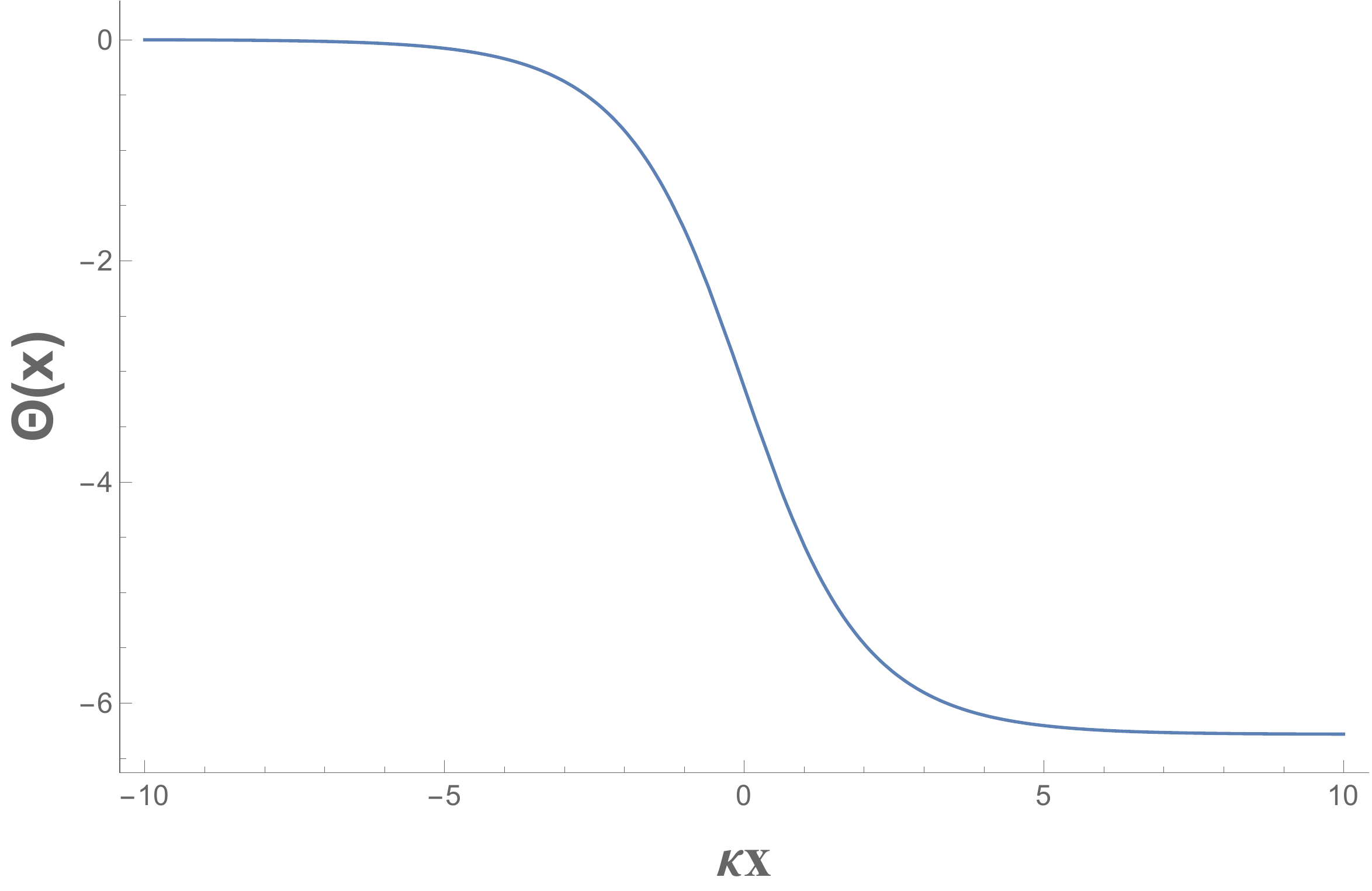}
\caption{This is the profile function $\Theta$ from Eq.~\eqref{eq:sg profile and magnetisation} plotted against the dimensionless variable $\kappa x$ for the critical value $\frac{B}{\kappa^{2}}=\frac{\pi^{2}}{16}$ at $A=0$.
}
\label{fig:anisotropy domain wall}
\end{center}
\end{figure}

\paragraph{$A=0, B>0$ Case :} 
One period of the spiral solution \eqref{eq: solution_ellipticFunc_A=0} 
reduces to two sets of the following domain wall solution in the 
limit of infinite period $k\to 1$ corresponding to 
$B=\frac{(\pi\kappa)^{2}}{16}$
%\begin{widetext}
\begin{equation}
\begin{split}
\Theta&=-4\arctan\left(\exp(\sqrt{B}x)\right),\\ 
\vec{n}^{\alpha}&=\left(0, 2\frac{\tanh\left(\sqrt{B}x\right)}
{\cosh\left(\sqrt{B}x\right)}, 
1-2\sech^{2}\left(\sqrt{B}x\right)\right)^{T}. 
\end{split}\label{eq:sg profile and magnetisation}
\end{equation}
%\end{widetext}
We note that this is a domain wall solution for any $B>0$ with 
$A=0$.
The total energy of the domain wall is zero at the phase transition 
point $B=\frac{(\pi\kappa)^{2}}{16}$, positive in the positively 
polarized ferromagnetic phase $B>\frac{(\pi\kappa)^{2}}{16}$, and 
negative in the spiral phase  $B<\frac{(\pi\kappa)^{2}}{16}$. 
We plot this on the right in
Fig.~\ref{fig:anisotropy domain wall}.
%\begin{figure}[htbp]
%\begin{center}
%\includegraphics[width=0.5\textwidth]{figures/sgdw2}
%\caption{The profile function $\Theta$ from Eq.~\eqref{eq:sg profile and magnetisation} plotted against the dimensionless variable $\kappa x$ for the critical value $\frac{B}{\kappa^{2}}=\frac{\pi^{2}}{16}$ at $A=0$.} 
%\label{fig:Sine-Gordon domain wall}
%\end{center}
%\end{figure}
It should be noted that while we have referred to the configurations as domain walls for $A=0, B>0$ they do not connect two different magnetic domains. In fact they interpolate between a magnetic domain and itself, since their phase rotates by a full $2\pi$. Some times these configurations would be called kinks, however, we would reserve that for when they are genuine soliton solutions with positive energy above the ferromagnetic phase. As such we stick to calling them domain walls.

\section{Domain walls in chiral magnets}\label{sec:exactdomainwalls}

\subsection{Exact domain wall solutions}
\label{sec:domain-wall}

Since the field equation~\eqref{eq:theta-EOM2} involves 
only the potential parameters $A, B$ and is independent of the DM 
interaction parameter $\kappa$, the shapes of domain wall solutions 
are identical to those of the double sine-Gordon model. 
For instance, Ref.~\cite{Condat1983} gave domain wall solutions 
for the double sine-Gordon model, but in an entirely different 
physical context. 
The functional form of our exact solutions agree with those in 
Ref.~\cite{Condat1983} after adjusting for the different conventions. 
However, our energy functional involves the DM term, which gives 
a crucial negative energy contribution for the energy of domain 
wall solution as solitons. 
This fact is the basis of our finding of zero energy domain 
wall solutions at the boundary between the homogeneous phase 
and the spiral phase. 
We will derive exact domain wall solutions and evaluate their energy, 
in order to be reasonably self-contained.

A domain wall is a soliton with a localized energy. 
We will demonstrate explicitly that it has 
positive energy in the homogeneous phase, negative energy in 
the spiral phase, and zero energy at the phase boundary. 
The domain wall solutions are found to correspond 
to the infinite period limit of the lowest energy spiral solution. 
These features show that domain wall solutions 
are solitons for excited states in the homogeneous phase, 
but are instability modes for the homogeneous background solution 
in the spiral phase. 
This picture agrees with the fact that the lowest energy spiral 
solution gives the ground state in the spiral phase region, 
which has lower average energy density than the homogeneous 
solution.

The domain wall solutions of our interest are defined as those 
solutions of the first order equation 
\eqref{eq: theta-equation} with the mechanical energy 
tuned as $C_0=-U_{\rm min}$. 
The first order equation for $\Theta(x)$ in Eq.~\eqref{eq: theta-equation} 
becomes 
\begin{equation}
\frac{d\Theta}{dx}=- \sqrt{2(-B\cos\Theta+A\cos^2\Theta+U_{\rm min})}, 
 \label{eq: theta-equation_DW}
\end{equation}
which gives a monotonically decreasing function $\Theta(x)$ 
connecting two adjacent minima $\Theta_{\rm min}$ of the 
potential.

\subsection{Domain wall solutions for \texorpdfstring{$0\le B \le 2A$}{0< B <2A}   }
\label{sec:domain-wall-Bless2A}

In the parameter region 
$0\le B \le 2A$, the minimum of the potential 
is $U_{\rm min}=-\frac{B^{2}}{4A}$, and the homogeneous phase is given 
by the canted ferromagnetic solution with $n_3=\frac{B}{2A}$. 
The first order field equation~\eqref{eq: theta-equation_DW} 
becomes 
\begin{equation}
\frac{d\Theta}{dx}=- \sqrt{2A}\left|\cos\Theta-\frac{B}{2A}\right|. 
 \label{eq: theta-equation_DW_Ble2A}
\end{equation}
Since there are two degenerate minima in one period $2\pi$, 
we have two different types of domain walls: 
the long wall connecting $\Theta=2\pi-\arccos\frac{B}{2A}$ at $x=-\infty$ 
to $\arccos\frac{B}{2A}$ at $x=\infty$, and the short wall connecting 
 $\Theta=\arccos\frac{B}{2A}$ at $x=-\infty$ 
to $-\arccos\frac{B}{2A}$ at $x=\infty$.

\subsubsection{Long wall }
\label{sec:long-wall}

\begin{figure}[htbp]
\begin{center}
\includegraphics[width=0.48\textwidth]{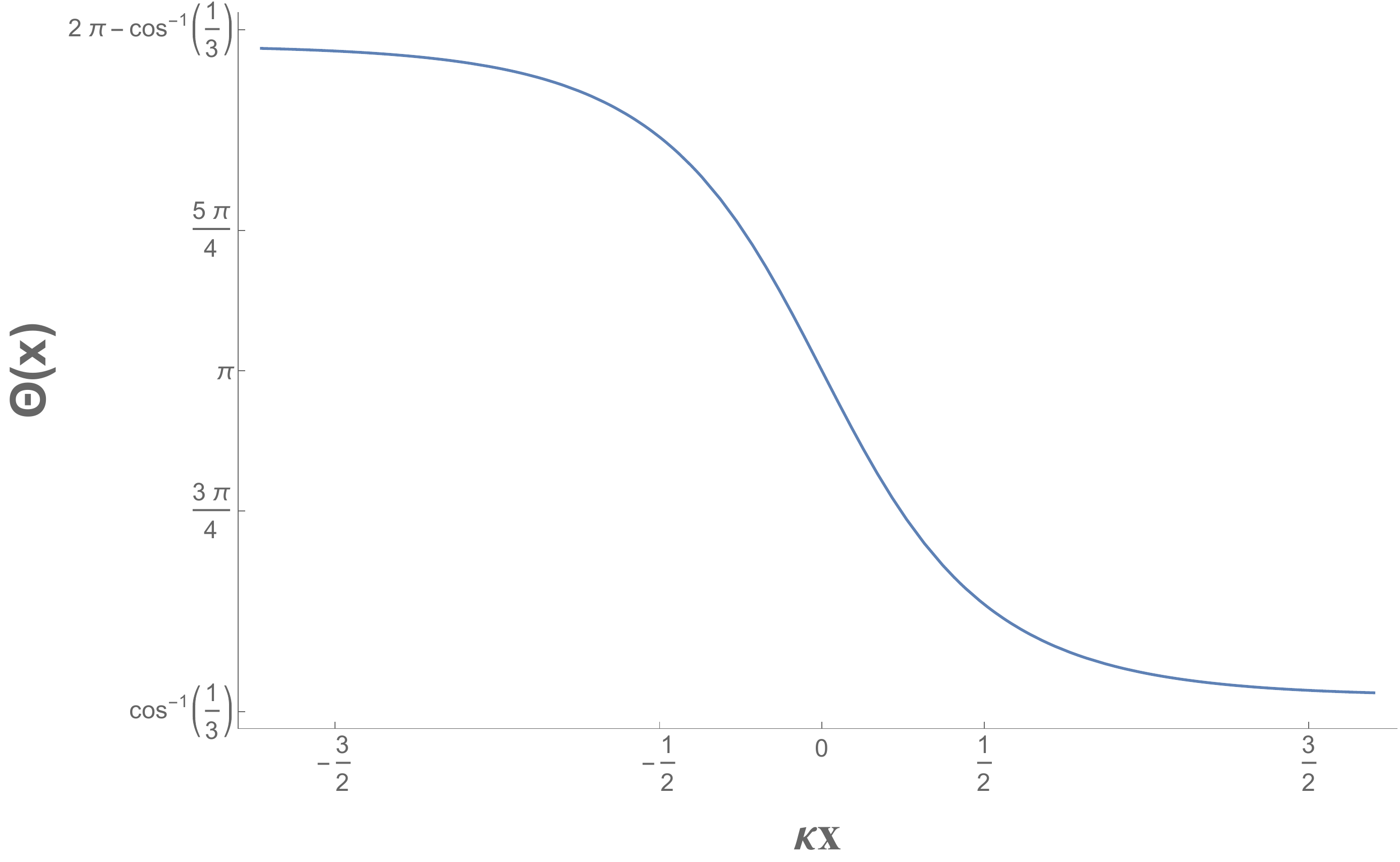}
\includegraphics[width=0.48\textwidth]{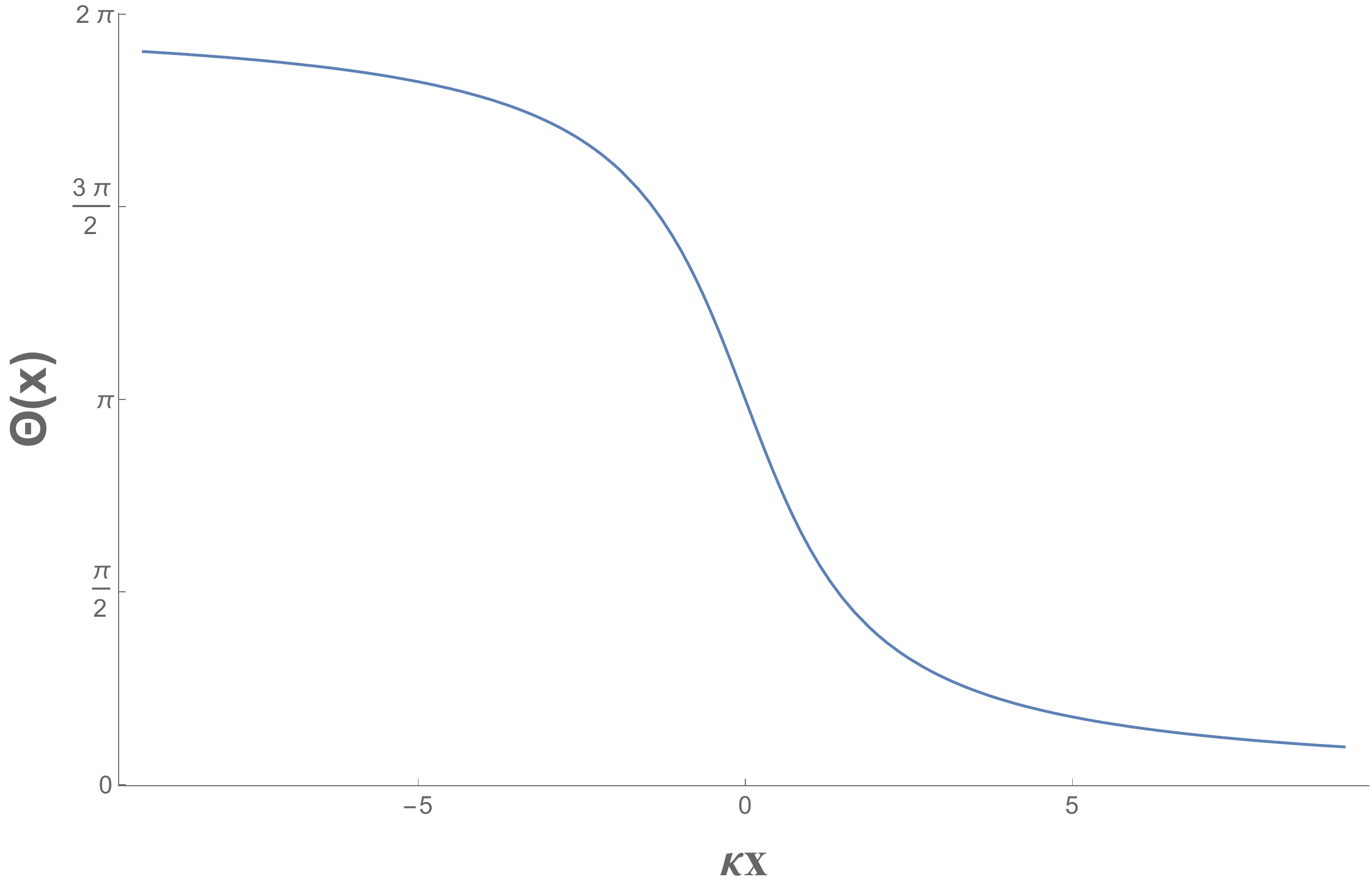}
\caption{On the left is a plot of the profile function $\Theta$ as a 
function of the dimensionless quantity $\kappa x$ for the long wall for $0\le B \le 2A$ in 
Eq.~\eqref{eq: long_DW_Bless2A}. 
We take $\frac{A}{\kappa^{2}}=3,\frac{B}{\kappa^{2}}=2$ as an example, and the 
domain wall to be centred at $x=0$. On the right is the profile of the domain wall solution $\Theta$ and in Eq.~\eqref{eq:DW_B=2A}
plotted against the dimensionless quantity $\kappa x$
for $B=2A=\frac{1}{2}\kappa^2$, with the domain wall to be centred at $x=0$.}
\label{fig:anisotropy dominated domain wall2}
\end{center}
\end{figure}

The long wall is obtained by integrating \eqref{eq: theta-equation_DW_Ble2A} 
with $\frac{B}{2A}-\cos\Theta\ge 0$. 
By choosing $\Theta=\pi$ at $x=0$ (midpoint of the domain wall), 
we obtain the solution connecting $\Theta=2\pi-\arccos\frac{B}{2A}$ 
at $x=-\infty$ to $\arccos\frac{B}{2A}$ at $x=\infty$ as 
%\begin{widetext}
\begin{align}
x&=
- \frac{1}{\sqrt{2A}}\int_\pi^\Theta 
\frac{d\Theta'}{\frac{B}{2A}-\cos\Theta'} 
\nonumber \\
&=\sqrt{\frac{2A}{4A^2-B^2}}\log\left(
\frac{\tan\frac{\Theta}{2}+\sqrt{\frac{2A-B}{2A+B}}}
{\tan\frac{\Theta}{2}-\sqrt{\frac{2A-B}{2A+B}}}\right)
\nonumber \\
&=\sqrt{\frac{2A}{4A^2-B^2}}\; 2\; {\rm arccoth}\left(
\sqrt{\frac{2A+B}{2A-B}}\tan\frac{\Theta}{2}\right)
. 
 \label{eq: long_DW_Bless2A}
\end{align}
%\end{widetext}
On the left in Fig.~\ref{fig:anisotropy dominated domain wall2}, we plot 
the domain wall profile $\Theta$ as a function of $x$.

In the limit of $2A \to B+0$, the domain wall solution reduces to 
\begin{eqnarray}
x= \frac{1}{\sqrt{B}}\cot\frac{\Theta}{2}, 
 \label{eq:DW_B=2A}
\end{eqnarray}
where we choose the branch $2\pi\ge\Theta\ge\pi$ for $-\infty\le x\le 0$, 
and $\pi\ge\Theta\ge0$ for $0\le x\le \infty$. 
This solution is depicted on the right in Fig.~\ref{fig:anisotropy dominated domain wall2}.

%\begin{figure}[htbp]
%\begin{center}
%\includegraphics[width=0.5\textwidth]{figures/BPS_dw_profile2}
%\caption{The profile of the domain wall solution $\Theta$ and in Eq.~\eqref{eq:DW_B=2A}
%plotted against the dimensionless quantity $\kappa x$
%for $B=2A=\frac{1}{2}\kappa^2$, with the domain wall to be centred at $x=0$.}
%\label{fig:BPS domain wall}
%\end{center}
%\end{figure}

\subsubsection{Short wall}
\label{sec:short-wall}

\begin{figure}[htbp]
\begin{center}
\includegraphics[width=0.5\textwidth]{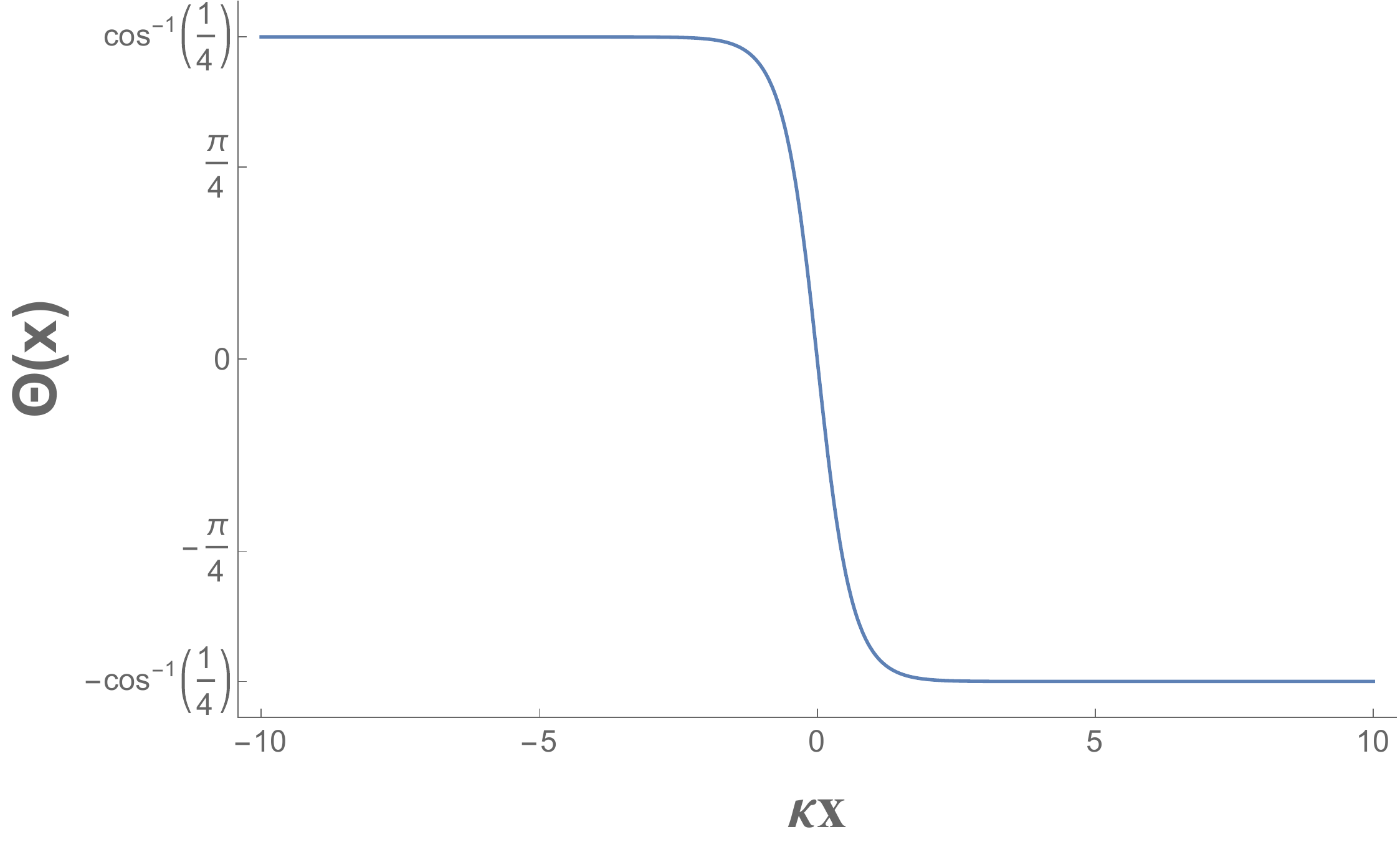}
\caption{This is a plot of the profile function $\Theta$ as a 
function of the dimensionless quantity $\kappa x$ for the short wall for $0\le B \le 2A$ in 
Eq.~\eqref{eq: short_DW_Bless2A}. 
We take $\frac{A}{\kappa^{2}}=4,\frac{B}{\kappa^{2}}=2$ as an example, and the domain wall to be 
centered at $\kappa x=0$. }
\label{fig:anisotropy dominated domain wall}
\end{center}
\end{figure}

The short wall is obtained by integrating \eqref{eq: theta-equation_DW_Ble2A} 
with $\cos\Theta-\frac{B}{2A}\ge 0$. 
By choosing $\Theta=0$ at $x=0$ (midpoint of the domain wall), 
we obtain the solution connecting $\Theta=\arccos\frac{B}{2A}$ 
at $x=-\infty$ to $-\arccos\frac{B}{2A}$ at $x=\infty$ as 
%\begin{widetext}
\begin{align}
x&=
- \frac{1}{\sqrt{2A}}\int_0^\Theta 
\frac{d\Theta'}{\cos\Theta'-\frac{B}{2A}} 
\nonumber \\
&=\sqrt{\frac{2A}{4A^2-B^2}}\log\left(
\frac{\sqrt{\frac{2A-B}{2A+B}}-\tan\frac{\Theta}{2}}
{\sqrt{\frac{2A-B}{2A+B}}+\tan\frac{\Theta}{2}}\right)
\nonumber \\
&=-\sqrt{\frac{2A}{4A^2-B^2}}\; 2\; {\rm arctanh}\left(
\sqrt{\frac{2A+B}{2A-B}}\tan\frac{\Theta}{2}\right)
. 
 \label{eq: short_DW_Bless2A}
\end{align}
%\end{widetext}
In Fig.~\ref{fig:anisotropy dominated domain wall}, we plot 
the domain wall profile $\Theta$ as a function of the dimensionless quantity $\kappa x$.

In the limit of $B\to0$, the long wall connects $\Theta=\frac{3\pi}{2}$ 
to $\frac{\pi}{2}$, whereas the short wall connects $\Theta=\frac{\pi}{2}$ to 
$-\frac{\pi}{2}$. 
We find that the long and short walls at $B\to 0$ are precisely 
identical to those walls emerging from the infinite period 
limit of the spiral solution at $B=0$. %in Sect.~\ref{sec:positiveA_spiral}. 

\subsubsection{Energy of long and short domain walls}
\label{sec:energy_long-short-wall}

For the long wall, the value of $\Theta$ spans from 
$2\pi-\arccos\frac{B}{2A}$ at $x=-\infty$ to $\arccos\frac{B}{2A}$ 
at $x=\infty$. 
The total energy of the long domain wall on the homogeneous 
solution (with energy density $U_{\rm min}$) as background is 
given by integrating over the energy density in 
Eq.~\eqref{eq:energy-density-solution} with $C_0=-U_{\rm min}$:
\begin{align}
{E}_{\rm long}&=\int_{-\infty}^{\infty}dx 
\left({\mathcal E}_{\rm sol}-U_{\rm min}\right)
\nonumber \\
&=\int_{-\infty}^{\infty}dx 
\left(\kappa\frac{d\Theta}{dx}
+2(U(\Theta)-U_{\rm min})\right) .
\label{eq:energy-long-wall1}
\end{align}
Changing variable from $x$ to $\Theta$, we obtain by using 
the domain wall equation in Eq.~\eqref{eq: theta-equation_DW} 
with $\frac{d\Theta}{dx}=-\sqrt{2(U(\Theta)-U_{\rm min})}
=-\sqrt{2A}(\frac{B}{2A}-\cos\Theta)$: 
%\begin{widetext}
\begin{align}
{E}_{\rm long}&=\int_{2\pi-\arccos\frac{B}{2A}}^{\arccos\frac{B}{2A}}
d\Theta \left(\frac{d\Theta}{dx}\right)^{-1}
\left(\kappa\frac{d\Theta}{dx}+2A\left(\frac{B}{2A}-\cos\Theta\right)^2
\right)
\label{eq:energy-long-wall2}
 \\
&=-2\kappa\left(\pi-\arccos\frac{B}{2A}\right)
+\frac{2B}{\sqrt{2A}}\left(\pi-\arccos\frac{B}{2A}\right)
+2\sqrt{2A-\frac{B^2}{2A}} .
\nonumber
\end{align}
%\end{widetext}
Similarly, the total energy of the short domain wall on the homogeneous 
solution as the background is given by integrating over the energy density. 
The only difference from the long wall is that $\Theta$ varies 
from $\arccos\frac{B}{2A}$ at $x=-\infty$ to $-\arccos\frac{B}{2A}$ 
at $x=\infty$, and $\cos\Theta-\frac{B}{2A}\ge0$. 
By noting $\frac{d\Theta}{dx}=-\sqrt{2(U(\Theta)-U_{\rm min})}
=-\sqrt{2A}(\cos\Theta-\frac{B}{2A})$, we find 
%\begin{widetext}
\begin{align}
{E}_{\rm short}&=\int_{-\arccos\frac{B}{2A}}^{\arccos\frac{B}{2A}}
d\Theta \left(\frac{d\Theta}{dx}\right)^{-1}
\left(\kappa\frac{d\Theta}{dx}+2A\left(\cos\Theta-\frac{B}{2A}\right)^2
\right)
\label{eq:energy-short-wall} \\
&=-2\kappa\arccos\frac{B}{2A}
-\frac{2B}{\sqrt{2A}}\arccos\frac{B}{2A}
+2\sqrt{2A-\frac{B^2}{2A}} .
\nonumber
\end{align}
%\end{widetext}
In order to obtain domain wall solutions as the infinite  period limit of a spiral solution, we need to consider domain walls 
connecting minimum of the potential which are $\Delta\Theta=2\pi$ apart. 
Such limiting domain walls precisely correspond to the sum of 
a pair of long and short domain walls. 
The sum of the energy of the short and long domain wall is given by 
%\begin{widetext}
\begin{align}
\label{eq:sum_energy-long-short}
E_{\rm period}&={E}_{\rm long}+E_{\rm short}
\\
&=-2\pi\kappa
+\frac{4B}{\sqrt{2A}}\left(\frac{\pi}{2}-\arccos\frac{B}{2A}\right)
+4\sqrt{2A-\frac{B^2}{2A}} .
\nonumber 
\end{align}
%\end{widetext}
Thus we see that the energy of a pair of long and short domain 
wall solutions is positive in the (homogeneous) ferromagnetic 
phase, negative in the spiral phase, and zero along the dashed red line in Fig.~\ref{fig:Phase boundaries}. 
Considering both the $B\to 2A$ and the $B\to 0$ limits of the long and short walls we see that they have the same energy in the $B\to 0$ limit, while the long wall has lower energy in the $B\to 2A$ limit. In fact for $B\to 2A$ the short domain wall disappears and has zero energy.

\subsection{Domain wall solutions for \texorpdfstring{$B>0$}{B>0} and \texorpdfstring{$B>2A$}{B>2A} }
\label{sec:domain-wall-2AlessB}

\begin{figure}[htbp]
\begin{center}
\includegraphics[width=0.5\textwidth]{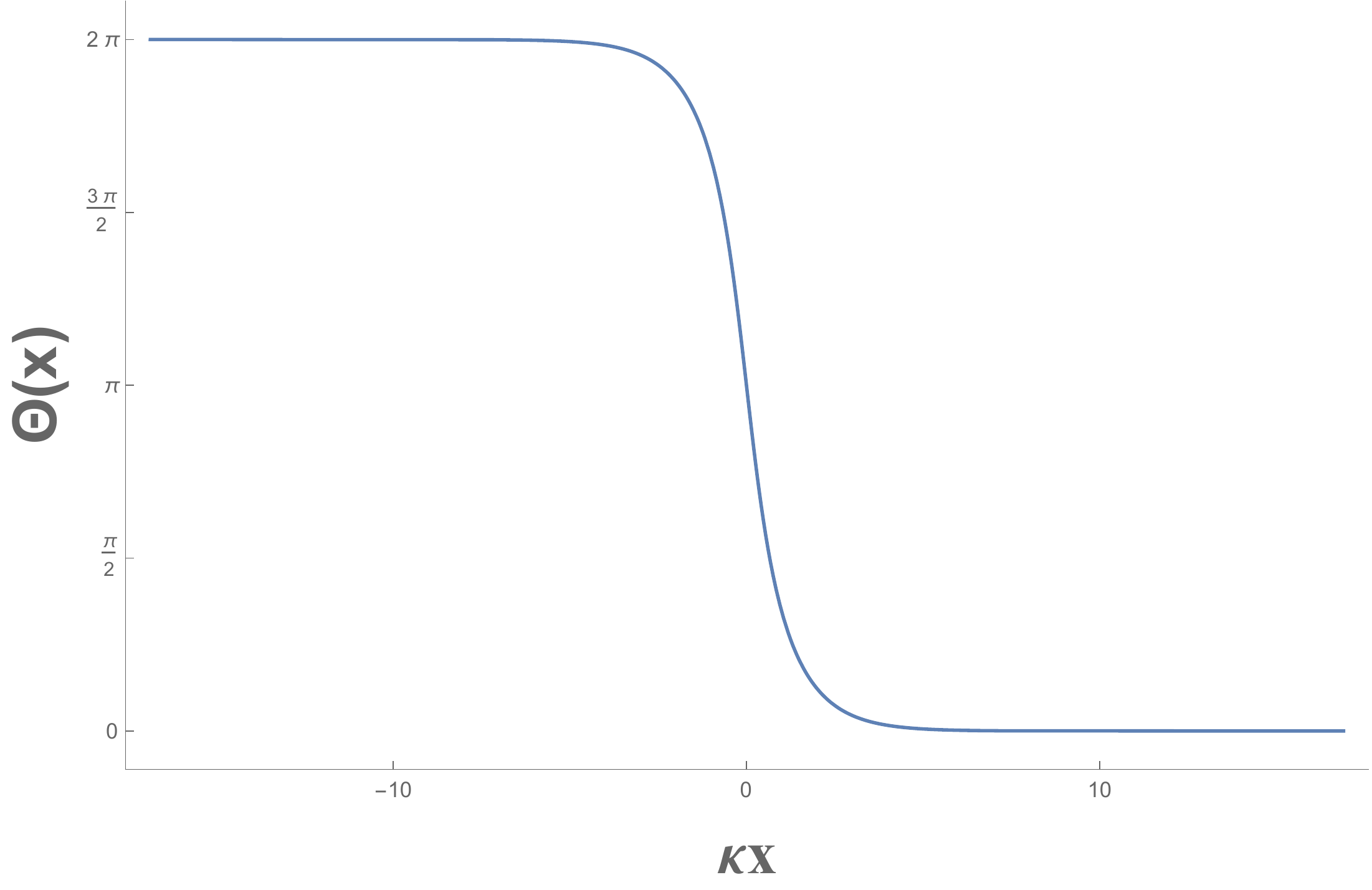}
\caption{This is a plot of the profile function $\Theta$ as a 
function of the dimensionless variable $\kappa x$ for a domain wall in the magnetic field dominated 
region $B\geq 2A$ in Eq.~\eqref{eq: DW_2AlessB}. 
We take $A=\frac{\kappa^2}{2},B=2\kappa^2$ as an example, and 
the domain wall to be 
centered at $x=0$. 
}
\label{fig:B dominated domain wall}
\end{center}
\end{figure}

In the parameter region 
$B>0$ and $B>2A$, the minimum of the potential 
is $U_{\rm min}=A-B$, and the homogeneous phase is given 
by the polarized ferromagnetic solution with $n_3=1$. 
The first order equation \eqref{eq: theta-equation_DW} 
becomes 
\begin{equation}
\frac{d\Theta}{dx}=- \sqrt{2(1-\cos\Theta)(B-A-A\cos\Theta)}. 
 \label{eq: theta-equation_DW_2AleB}
\end{equation}
In the region 
$B>0$ and $B>2A$, the minimum of the potential occur at 
$\Theta=2\pi{\mathbb Z}$. 
Let us consider a domain wall connecting $\Theta=2\pi$ at $x=-\infty$ 
to $\Theta=0$ at $x=\infty$. 
By choosing the midpoint of the domain wall $\Theta=\pi$ at $x=0$, we find 
%\begin{widetext}
\begin{align}
x&=
- \int_\pi^\Theta 
\frac{d\Theta'}{\sqrt{2(1-\cos\Theta')(B-A-A\cos\Theta')}} 
\nonumber \\
&=\frac{1}{\sqrt{B-2A}}\log\left(
\sqrt{1+\left(1-\frac{2A}{B}\right)\cot^2\frac{\Theta}{2}}
+\sqrt{1-\frac{2A}{B}}\cot\frac{\Theta}{2}\right). 
 \label{eq: DW_2AlessB}
\end{align}
%\end{widetext}
In Fig.~\ref{fig:B dominated domain wall}, we plot 
the domain wall profile $\Theta$ as a function of $x$. 
In the limit of $2A \to B-0$, the domain wall solution reduces to 
an identical form to the domain wall obtained in the limit 
of $2A\to B+0$ 
from Eq.~\eqref{eq:DW_B=2A}.

Since the energy density of the homogeneous solution (polarized 
ferromagnetic ground state) is $U_{\rm min}=A-B$, the total energy 
of the domain wall on the homogeneous background is given by 
%\begin{widetext}
\begin{align}
{E}_{\rm DW}
&=\int_{-\infty}^{\infty}dx 
\left(\kappa\frac{d\Theta}{dx}+2(U(\Theta)-U_{\rm min})\right)
\nonumber \\
%&=&\int_{2\pi}^{0}d\Theta \left(\frac{d\Theta}{dx}\right)^{-1}\left(\kappa\frac{d\Theta}{dx}+2(1-\cos\Theta)(B-A-A\cos^2\Theta)\right)\nonumber \\
&=
-2\pi\kappa+4\sqrt{B-2A}
+\frac{4B}{\sqrt{2A}}\arcsin\sqrt{\frac{2A}{B}},
\label{eq:energy-domain-wall_2AleB}
\end{align}
%\end{widetext}
in the case of $A> 0$, and 
%\begin{widetext}
\begin{equation}
{E}_{\rm DW}
=
-2\pi\kappa+4\sqrt{B-2A}
+\frac{4B}{\sqrt{-2A}}\arcsinh\sqrt{\frac{-2A}{B}}, 
\label{eq:energy-domain-wall_2AleB-2}
\end{equation}
%\end{widetext}
in the case of $A\le0$. 
We see that the zero energy condition for the domain wall gives 
the boundary between the homogeneous phase and spiral phase, 
as given in Eqs.~\eqref{eq:phase-boundary-case2} and 
\eqref{eq:phase-boundary-case2-2}. 
Moreover, the domain wall is a positive energy soliton as an 
excited state in the homogeneous phase. 
In the spiral phase, on the other hand, 
the domain wall on the (homogeneous) ferromagnetic background 
solution gives a negative energy solution, signaling the instability 
of the ferromagnetic (homogeneous) solution. 
These observations confirm our conclusion that the homogeneous 
solution is unstable in the spiral phase and decays into the 
spiral solution, giving the ground state.

\section{Conclusion and discussion }\label{sec:summary}

We have studied solutions of the field equations 
for chiral magnets in one spatial dimension, 
in order to determine their phase diagram. 
There are three distinct homogeneous phases: the canted polarized 
ferromagnetic phase with $n_3\not=\pm 1$, the positive polarized 
ferromagnetic phase with $n_3=1$, and the negative polarized phase 
with $n_3=-1$. 
These homogeneous ferromagnetic phases are realized for large 
values of potential parameters, whereas the spiral phase is 
realized for small values of the potential parameters. 
%In the spiral phase, there is a continuum of spiral solutions 
%among which the configuration with the lowest average energy density gives 
%the ground state. 
We have explicitly determined the exact phase boundary between spiral and homogeneous 
ferromagnetic phases. As mentioned previously, there is numerical evidence in \cite{Chovan} that a lower energy non-flat spiral state is possible when $B=0$. This state cannot be studied analytically which is why we have not discussed it here, and we have focused on the parameter region $2A\leq \vert B\vert$ where the flat spiral is the ground state.
The lowest energy spiral solution has an infinite period 
at the phase boundary, and becomes a domain wall solution with 
zero energy. 
We have constructed domain wall solutions and found that they have 
positive energy in the homogeneous ferromagnetic phases and 
have negative energy in the spiral phase, exhibiting the instability 
of the homogeneous ferromagnetic solution in the spiral phase region. 
We have also found that the phase transitions between spiral phase and 
homogeneous ferromagnetic phases are of second order. 

Our results should be useful when discussing the phase diagram of 
chiral magnets in spatial dimensions two or higher~\cite{BY,BH,
BRS,Schroers1,RSN,DM,Melcher}. 
The homogeneous and spiral solutions are also present 
in higher spatial dimensions. 
However, We need to examine possible instabilities due to the fluctuations 
around these solutions, in particular those emerging from additional 
dimensions. 
More importantly, we need to consider other inhomogeneous solutions 
that are intrinsically two or higher dimensional, such as 
skyrmions or merons and their lattices. 

Another interesting question to be addressed is the low energy 
effective field theory on the spiral ground state. 
Since the spiral solutions are spatially inhomogeneous, the 
low energy effective field theory is not invariant under translation 
nor rotation. Therefore the dispersion relation of fluctuations 
is  expected to exhibit asymmetry in this regards. 
It has been already computed for the simplest spiral ground state 
for the chiral magnet without the potential, and an interesting 
anisotropic dispersion relation was found in Ref.~\cite{Hongo:2020xaw}. 
Moreover, the frequency of the fluctuations depends on the 
structure of the time derivative, which is different for
ferromagnetic, antiferromagnetic, or ferrimagnetic materials  
\cite{Kobayashi:2015pra}.

For a single domain wall, the first order time derivative 
introduces a type-B NG mode due to
spontaneously broken translational and U(1) symmetries 
\cite{Kobayashi:2014xua}.

It is also interesting to consider a more general potential. 
The more general interaction terms arising from antiferromagnetic 
materials have been considered phenomenologically \cite{BRWM}. 

In this work, we considered only the effect of various solitonic 
objects such as domain walls and spiral solutions
in a 
mean-field approximation. 
It is an interesting future task to consider quantum effects, 
including nonperturbative effects in the chiral magnet.

%%%%%%%%%%%%%%%%%%%%%%%%%%%%%%%%%%%%%%%%%%%%%%%%%%%%%
%%%%%%%%%%%%%%%%%%%%%%%%%%%%%%%%%%%%%%%%%%%%%%%%%%%%%
\begin{acknowledgments}
This work is supported in part by the Japan Society for the Promotion 
of Science (JSPS) 
Grant-in-Aid for Scientific Research (KAKENHI) Grant Number 
18H01217 (M.~N. and N.~S.). 
\end{acknowledgments}

%%%%%%%%%%%%%%%%%%%%%%%%%%%%%%%%%%%%%%%%%%%%%%%%%%%%%
%%%%%%%%%%%%%%%%%%%%%%%%%%%%%%%%%%%%%%%%%%%%%%%%%%%%%
\appendix

\section{Derivative of energy density of spiral solutions} 
\label{sec:der_spiral}

To obtain the second derivative of energy density of the lowest 
energy spiral solution, we differentiate Eqs.~\eqref{eq:1st-derivative-B}
and \eqref{eq:1st-derivative-A} in terms of $B, A$. 
We can express them in terms of the weighted integrals defined in 
Eq.~\eqref{eq:weighted-integral} as 
\begin{equation}
L \frac{\partial^2 \Delta C}{\partial A^2}
=(L_0^{01})^2\frac{L_1^{00}}{L^2}-2L_0^{01}\frac{L_1^{01}}{L}
+L_1^{02}, 
\label{eq:C-AA}
\end{equation}
\begin{equation}
L \frac{\partial^2 \Delta C}{\partial A \partial B}
=(L_0^{10}L_0^{01})\frac{L_1^{00}}{L^2}
-L_0^{10}\frac{L_1^{01}}{L}
-L_0^{01}\frac{L_1^{10}}{L}
+L_1^{11}, 
\label{eq:C-BA}
\end{equation}
\begin{equation}
L \frac{\partial^2 \Delta C}{\partial B^2}
=(L_0^{10})^2\frac{L_1^{00}}{L^2}-2L_0^{10}\frac{L_1^{10}}{L}
+L_1^{20}, 
\label{eq:C-BB}
\end{equation}
We need to evaluate them at the phase boundary, namely in the 
limit of $\Delta C \to 0$. 
In the limit, the period $L$ diverges logarithmically as 
$\log(\frac{1}{\Delta C})$ because of the singularity of the integrand at 
$\Theta_{\rm min}$. 
More generally the weighted integral $L_n^{kl}$ diverges 
logarithmically $L_n^{kl} \sim \log(\frac{1}{\Delta C})$ if and only if 
$n=k+l$. 
If $n\ge k+l+1$, $L_n^{kl}$ diverges as $(\frac{1}{\Delta C})^{n-k-l}$, 
whereas they are finite if $n\le k+l-1$. 
The second derivative has only nonvanishing contributions from 
the power divergent term (the first term) because of $L\to \infty$ 
and becomes 
\begin{equation}
 \frac{\partial^2 \Delta C}{\partial A^2}
\sim(L_0^{01})^2\frac{L_1^{00}}{L^3}, 
\label{eq:C-AA2}
\end{equation}
\begin{equation}
 \frac{\partial^2 \Delta C}{\partial A \partial B}
\sim (L_0^{10}L_0^{01})\frac{L_1^{00}}{L^3}
, 
\label{eq:C-BA2}
\end{equation}
\begin{equation}
\frac{\partial^2 \Delta C}{\partial B^2}
\sim (L_0^{10})^2\frac{L_1^{00}}{L^3}
, 
\label{eq:C-BB2}
\end{equation}
These give Eq.~\eqref{eq:C-der-bound}.

We can evaluate the leading behavior of $L_n^{kl}$ in the limit 
of $\Delta C\to 0$ more explicitly by distinguishing the three 
parameter regions I, II, and III, since the values $U_{\rm min}$ 
and the associated $\Theta_{\min}$ are different in three regions.
Let us assume $B\ge 0$, since $B\le 0$ case can be obtained 
by using the symmetry $n_3\to -n_3, B \to -B$.

In region I, we have $B\le 2A$ and $\Theta_{\rm min}=\arccos \frac{B}{2A}$, 
and $U_{\rm min}=-\frac{B^2}{4A}$. 
In this case, the weighted integral becomes using $t=\cos\Theta$ 
%\begin{widetext}
\begin{align}
\label{eq:weighted-integral_BleA}
L_n^{kl}&=\int_0^{2\pi}d\Theta
\frac{\left(\frac{B}{2A}-\cos\Theta\right)^k
\left(\cos^2\Theta-\left(\frac{B}{2A}\right)^2\right)^l}
{\left[2\left(A\left(\cos\Theta-\frac{B}{2A}\right)^2
+\Delta C\right)\right]^{n+1/2}}
\\
&=2\int_{-1}^{1}\frac{dt}{\sqrt{1-t^2}}
\frac{\left(\frac{B}{2A}-t\right)^k
\left(t^2-\left(\frac{B}{2A}\right)^2\right)^l}
{\left[2\left(A\left(t-\frac{B}{2A}\right)^2
+\Delta C\right)\right]^{n+1/2}}\nonumber
. 
\end{align}
%\end{widetext}
For $n\le k+l-1$, $L_n^{kl}$ are finite in the limit of $\Delta C\to 0$. 
For $n=k+l$, $L_n^{kl}$ are logarithmically divergent as $\Delta C\to 0$. 
We find that the leading behavior is given by 
\begin{equation}
L_{n=k+l}^{kl}=\frac{2\sqrt{2A}}{\sqrt{4A^2-B^2}}
\left(\frac{-1}{2}\right)^k
\left(\frac{B}{2A^2}\right)^l 
\left(\log\frac{1}{\Delta C}+{\rm constant}\right)
. 
\label{eq:log-div-weighted-integral_BleA}
\end{equation}
For $n\ge k+l+1$, $L_n^{kl}$ are divergent in powers of $\frac{1}{\Delta C}$. 
We find that the leading behavior is given by 
%\begin{widetext}
\begin{align}
L_{n}^{kl}&=\frac{4\sqrt{2A}}{\sqrt{4A^2-B^2}}\frac{1}{2^n}
\left(\frac{-1}{A}\right)^k
\left(\frac{B}{A^2}\right)^l 
\frac{1}{(\Delta C)^{n-k-l}}\sum_{r=0}^{n-k-l-1}\frac{(-1)^r}{2r-1}\left(
\begin{array}{c}
      n-k-l-1 \\
      r
    \end{array}
\right)
. \label{eq:power-div-weighted-integral_BleA}
\end{align}
%\end{widetext}

In region II, we have $B\ge 2A$ and $\Theta_{\rm min}=2\pi {\mathbb Z}$, 
and $U_{\rm min}=A-B$. 
In this case, the weighted integral becomes using $t=\cos\Theta$ 
%\begin{widetext}
\begin{align}
\label{eq:weighted-integral_BgeA}
L_n^{kl}&=\int_0^{2\pi}d\Theta
\frac{\left(1-\cos\Theta\right)^k
\left(\cos^2\Theta-1\right)^l}
{\left[2\left((1-\cos\Theta)(B-A-A\cos\Theta)
+\Delta C\right)\right]^{n+1/2}}\nonumber
\\
&=2\int_{-1}^{1}\frac{dt}{\sqrt{1-t^2}}
\frac{\left(1-t\right)^k
\left(t^2-1\right)^l}
{\left[2\left((1-t)(B-A-At)
+\Delta C\right)\right]^{n+1/2}}
. 
\end{align}
%\end{widetext}
For $n\le k+l-1$, $L_n^{kl}$ are finite in the limit of $\Delta C\to 0$. 
For $n=k+l$, $L_n^{kl}$ are logarithmically divergent as $\Delta C\to 0$. 
We find that the leading behavior is given by 
\begin{equation}
L_{n=k+l}^{kl}=\frac{(-1)^l}{2^k(B-2A)^{k+l+\frac{1}{2}}}
\left(\log\frac{1}{\Delta C}+{\rm constant}\right)
. 
\label{eq:log-div-weighted-integral_BgeA}
\end{equation}
For $n\ge k+l+1$, $L_n^{kl}$ are divergent in powers of $\frac{1}{\Delta C}$. 
We find that the leading behavior is given by for $n\ge k+l+1\ge 2$ 
\begin{equation}
L_{n}^{kl}=\frac{(-1)^l}{2^{n-l}(B-2A)^n}\frac{1}{(B-2A)^{k+l+\frac{1}{2}}}
\frac{1}{(\Delta C)^{n-k-l}}
. 
\label{eq:power-div-weighted-integral_BgeA}
\end{equation}
For $n\ge 1, k=l=0$, we find that the leading behavior is given by
\begin{equation}
L_{n}^{00}=\frac{1}{2^{n-1}\sqrt{B-2A}}
\frac{1}{(\Delta C)^{n}}
\sum_{r=0}^{n-1}\frac{(-1)^r}{2r+1}
\left(
\begin{array}{c}
      n-1 \\
      r
    \end{array}
\right)
. 
\label{eq:power-div-weighted-integral_BgeA2}
\end{equation}

\section{Exact spiral solutions for \texorpdfstring{$B=0$}{B=0} or \texorpdfstring{$A=0$}{A=0}}
\label{sec:exact_sol_AorB}
Here we give some details about the spiral solutions in Section.~\ref{sec:exact_spiral_solution}. These are standard results about elliptic functions and some of them can also be found in the discussion of the spiral configurations in \cite{BH,Chovan}.
 \subsection{Spirals without external magnetic fields (\texorpdfstring{$B=0$}{B=0})}
\label{app:positiveA_spiral}
\subsubsection{Case of positive anisotropy (\texorpdfstring{$A>0$}{A>0})}
In the case of $A>0, B=0$, we obtain the flat spiral solution as 
\begin{equation}
\sqrt{2(C_0+A)}(x-x_0)=-\int_{{\Theta}(x_0)}^{{\Theta}(x)} 
\frac{d{\Theta}}{\sqrt{1-k^2\sin^2{\Theta}}} \; , 
 \label{eq: theta-solution_B=0+}
\end{equation}
where the modulus parameter of the elliptic integral $k$ 
is given by 
\begin{equation}
k=\sqrt{\frac{A}{C_0+A}}. 
\label{eq:elliptic_k_B=0+}
\end{equation}
As mentioned in the main body, these are not the lowest energy solutions as a non-flat spiral exists which can only be studied numerically. The flat spirals still exist, just not as the ground state, so we include details about them for completeness.
We find that $k<1$ for $C_0>-U_{\rm min}=0$, implying 
monotonically decreasing solutions. We can express 
the solution in terms of a Jacobi elliptic function~\cite{gradshtein2007} 
(choosing the NG mode to be $\Theta(x_0=0)=0$) as 
\begin{equation}
\sin{\Theta}=-{\rm sn}\left(\sqrt{2(C_0+A)}x, k\right) ,
 \label{eq: solution_ellipticFunc_B=0+}
\end{equation}
where the Jacobi elliptic function ${\rm sn}(u,k)$ is defined 
by $\sin \phi={\rm sn}(u,k)$ as the inverse function of the 
integral 
\begin{equation}
u=\int_0^\phi \frac{d\varphi}
{\sqrt{1-k^2\sin^2\varphi}} .
 \label{eq: elliptic_sn}
\end{equation}
The period 
$L$ and the average energy density $\langle \mathcal{E}\rangle$ 
of the spiral solution are given as 
\begin{equation}
L=\frac{4}{\sqrt{2(C_0+A)}} K(k) ,
 %\label{eq: period_B=0+}
\end{equation}
\begin{equation}
\langle \mathcal{E}\rangle =\frac{f(C_0;A>0,B=0)}{L}-C_0 ,
 %\label{eq: average_energy_density_B=0+}
\end{equation}
respectively. The excess energy defined in Eq.~\eqref{eq:f-no-pot} 
becomes 
\begin{equation}
f(C_0;A>0,B=0)=-2\pi\kappa 
+4\sqrt{2(C_0+A)}E(k) .
% \label{eq: excess_energy_B=0+}
\end{equation}
The lowest energy flat spiral solution occurs when $f(C_0;A,B=0)=0$, 
giving 
\begin{equation}
A=\frac{\pi^{2}\kappa^{2}}{8}\left(\frac{k}{E(k)}\right)^2, 
\label{eq:minimum_cond_B=0+}
\end{equation}
which determines $k$ in terms of $A$. 
Combining with Eq.~\eqref{eq:elliptic_k_B=0+}, we find $C_0(A)$ 
also as a function of $A$.
 
For $B\neq 0$ but $B\leq 2A$ the excess energy vanishes when
\begin{align}
2\pi\kappa&=\int_0^{2\pi}d\Theta
\sqrt{2\left(-B\cos\Theta+A\cos^2\Theta +\frac{B^2}{4A}\right)}
\nonumber \\
&=4\sqrt{2A-\frac{B^2}{2A}}
+2\sqrt{2A}\frac{B}{A}\left(\frac{\pi}{2}-\arccos\frac{B}{2A}\right). 
\label{eq:phase-boundary-case1}
\end{align} 
This is not a phase boundary since the spiral is not a ground state, it is merely where the flat spiral has zero energy. This is the dashed red curve included in Fig.~\ref{fig:Phase boundaries}.

\begin{figure}[htbp]
\begin{center}
\includegraphics[width=0.5\textwidth]{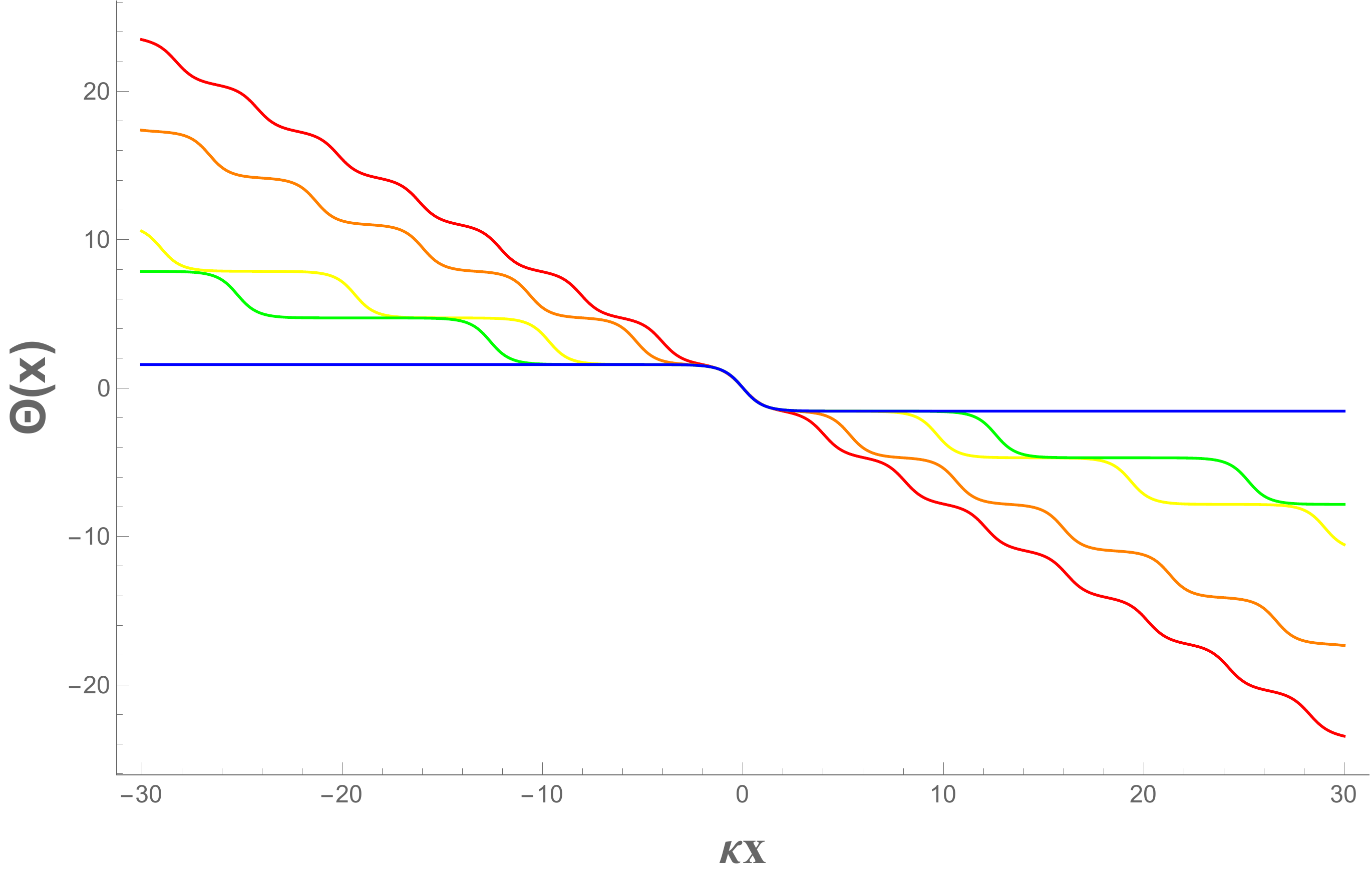}
\caption{These are plots of the profile 
$\Theta(x)$ for the lowest energy spiral solution, $A$ given by Eq.~\eqref{eq:minimum_cond_B=0+}, at  several values of $k$: Red $k=\frac{98}{100}$, 
Orange $k=\frac{998}{100}$, Yellow $k=\frac{999998}{1000000}$, 
Green $k=\frac{99999998}{100000000}$, and Blue $k=1$. As the 
modulus $k$ increases $\Theta(x)$ tends towards a domain wall 
configuration, achieving it for $k=1$.
}
\label{fig:anisotropy_spiral_state}
\end{center}
\end{figure}

Examples of the profile function for different values of the modulus 
$k$ are given in Fig.~\ref{fig:anisotropy_spiral_state}.
As can be seen in Fig.~\ref{fig:anisotropy_spiral_state}, the 
domain wall solution connects $\Theta=\frac{\pi}{2}$ at $x=-\infty$ to 
$\Theta=-\frac{\pi}{2}$ at $x=\infty$. 
Since one period of the spiral solution should span $\Delta\Theta=-2\pi$, 
the infinite period limit of the one period of spiral solution contains 
two domain wall solutions. These are the long and short domain wall configurations of Equations.~\eqref{eq: long_DW_Bless2A} and \eqref{eq: short_DW_Bless2A}.

One period of the spiral solution \eqref{eq: solution_ellipticFunc_B=0+} 
reduces to the following domain wall solution in the limit of infinite 
period $k\to 1$ corresponding to $A=\frac{(\pi\kappa)^{2}}{8}$
%\begin{widetext}
\begin{equation}
\begin{split}
\Theta&=-2\arctan\left(\exp(\sqrt{2A}x)\right)+\frac{\pi}{2} ,\\ \vec{n}^{\alpha}&=\left(0,-\tanh\left(%\frac{\pi}{2}
\sqrt{2A}x\right),
\sech\left(\sqrt{2A}x\right)\right)^{T}. 
\end{split}
\label{eq:anisotropy dw profile and magnetisation}
\end{equation}
%\end{widetext}
We note that this is a domain wall solution for any $A>0$ with 
$B=0$.
The total energy of the domain wall is zero at $A=\frac{(\pi\kappa)^{2}}{8}$, positive for $A>\frac{(\pi\kappa)^{2}}{8}$, and negative in 
the spiral phase  $A<\frac{(\pi\kappa)^{2}}{8}$.

\subsubsection{Case of negative anisotropy (\texorpdfstring{$A<0$}{A<0})}

This is formally the same as the previous case. However, the elliptic modulus is changed which modifies some of the results.
In the case of $A<0, B=0$, we define $\hat \Theta=\Theta+\frac{\pi}{2}$ 
to obtain 
\begin{equation}
\sqrt{2C_0}(x-x_0)=-\int_{{\hat\Theta}(x_0)}^{{\hat\Theta}(x)} 
\frac{d{\hat\Theta}}{\sqrt{1-k^2\sin^2{\hat\Theta}}} \; , 
 \label{eq: theta-solution_B=0-}
\end{equation}
where the modulus of the elliptic integral $k$ is given by 
\begin{equation}
k=\sqrt{\frac{-A}{C_0}}, 
%\label{eq:elliptic_k_B=0-}
\end{equation}
and the solution is given by means of Jacobi elliptic function as 
\begin{equation}
\cos\Theta=\sin{\hat\Theta}=-{\rm sn}\left(\sqrt{2C_0}x, k\right) ,
 \label{eq: solution_ellipticFunc_B=0-}
\end{equation}
where the NG mode is chosen as $\Theta(x_0=0)=\frac{\pi}{2}$. To find the expressions in this case simply take the expressions for the Period and energy density in the $A>0$ case and replace $C_{0}+A$ by $C_{0}$. Doing this we find
% The period 
%$L$ and the average energy density $\langle \mathcal{E}\rangle$ 
%of the spiral solution are given as 
%\begin{equation}
%L=\frac{4}{\sqrt{2C_0}} K(k) ,
% %\label{eq: period_B=0-}
%\end{equation}
%\begin{equation}
%\langle \mathcal{E}\rangle =\frac{f(C_0;A<0,B=0)}{L}-C_0 ,
% %\label{eq: average_energy_density_B=0-}
%\end{equation}
%respectively. The excess energy defined in Eq.~\eqref{eq:f-no-pot} 
%becomes 
%\begin{equation}
%f(C_0;A<0,B=0)=-2\pi\kappa 
%+4\sqrt{2C_0}E(k) .
% %\label{eq: excess_energy_B=0-}
%\end{equation}
that the lowest energy spiral solution occurs when $f(C_0;A<0,B=0)=0$, 
giving 
\begin{equation}
A=-\frac{\pi^{2}\kappa^{2}}{8}\left(\frac{k}{E(k)}\right)^2. 
\label{eq:minimum_cond_B=0-}
\end{equation}

Combining with Eq.~\eqref{eq:elliptic_k_B=0-}, we find $C_0(A)$ 
as a function of $A$, which determines $C_0(A)$ as a function of $A$. 
The minimum energy is given by 
\begin{equation}
\langle \mathcal{E}\rangle
 =-C_0(A). 
\label{eq:lowest_energy_B=0-}
\end{equation}
As $C_0$ approaches the lowest allowed value $-U_{\rm min}=A<0$, 
the period $L$ becomes infinite, $f(C_0=0;A<0,B=0)=0$, and the 
lowest energy spiral solution has the same average energy 
density as the homogeneous solution. 
This phase boundary of spiral and homogeneous phases occurs at 
the $C_0\to A$ limit of the minimum energy condition 
\eqref{eq:minimum_cond_B=0-} as 
\begin{equation}
A=-\frac{\pi^{2}}{8}\kappa^{2}. \label{eq:critical_anisotropy-}
\end{equation}

The domain wall solution in this case connects $\Theta=\pi$ at 
$x=-\infty$ to $\Theta=0$ at $x=\infty$. 
Since one period of the spiral solution should span $\Delta\Theta=-2\pi$, 
the infinite period limit of one period of the spiral 
solution contains two domain wall solutions. These are a long domain wall plus a short domain wall.

\subsection{Spiral solutions without anisotropy (\texorpdfstring{$A=0$}{A=0})}
Eq.~\eqref{eq: theta-equation_A=0} is solved using 
$\tilde{\Theta}=\frac{\Theta+\pi}{2}$ 
\begin{equation}
\sqrt{\frac{C_0+B}{2}}(x-x_0)=-\int_{\tilde{\Theta}(x_0)}^{\tilde{\Theta}(x)} 
\frac{d\tilde{\Theta}}{\sqrt{1-k^2\sin^2\tilde{\Theta}}} \; , 
 \label{eq: theta-solution_A=0}
\end{equation}
where the modulus of the elliptic integral $k$ is given by 
\begin{equation}
k=\sqrt{\frac{2B}{C_0+B}}. 
%\label{eq:elliptic_k_A=0}
\end{equation}
We find that $k<1$ for $C_0>-U_{\rm min}=B$. 
The monotonically decreasing solutions 
can be expressed in terms of a Jacobi elliptic 
function~\cite{gradshtein2007} as 
\begin{equation}
\cos\frac{\Theta}{2}=-{\rm sn}\left(\sqrt{\frac{C_0+B}{2}}x, k\right) ,
 \label{eq: solution_ellipticFunc_A=0}
\end{equation}
where the NG mode is chosen such that $\Theta(x_0=0)=-\pi$. 
The period $L$, the excess energy $f(C_0;A=0,B)$ in a period, 
and the average energy density 
$\langle \mathcal{E}\rangle$ of the spiral solution are given as 
\begin{equation}
L=2\sqrt{\frac{2}{C_0+B}} K(k) ,
 %\label{eq: period_A=0}
\end{equation}
\begin{equation}
f(C_0;A=0,B)=-2\pi\kappa 
+4\sqrt{2(C_0+B)}E(k) ,
 %\label{eq: excess_energy_A=0}
\end{equation}
\begin{equation}
\langle \mathcal{E}\rangle =\frac{f(C_0,A=0,B)}{L}-C_0 ,
 %\label{eq: average_energy_density_A=0}
\end{equation}
respectively. 
The lowest energy spiral solution occurs when $f(C_0;A=0,B)=0$, 
giving 
\begin{equation}
B=\frac{\pi^{2}\kappa^{2}}{16}\left(\frac{k}{E(k)}\right)^2, 
\label{SG k minimum}
\end{equation}
which determines $C_0(B)$ as a function of $B$. 
The minimum energy is given by 
\begin{equation}
\langle \mathcal{E}\rangle%_{\rm min}
 =-C_0(B), 
 %\label{eq:lowest_energy_spiral_A=0}
\end{equation}
which is lower than 
the positively polarized ferromagnetic state with the 
energy density $-B$.

\begin{figure}[htbp]
\begin{center}
\includegraphics[width=0.5\textwidth]{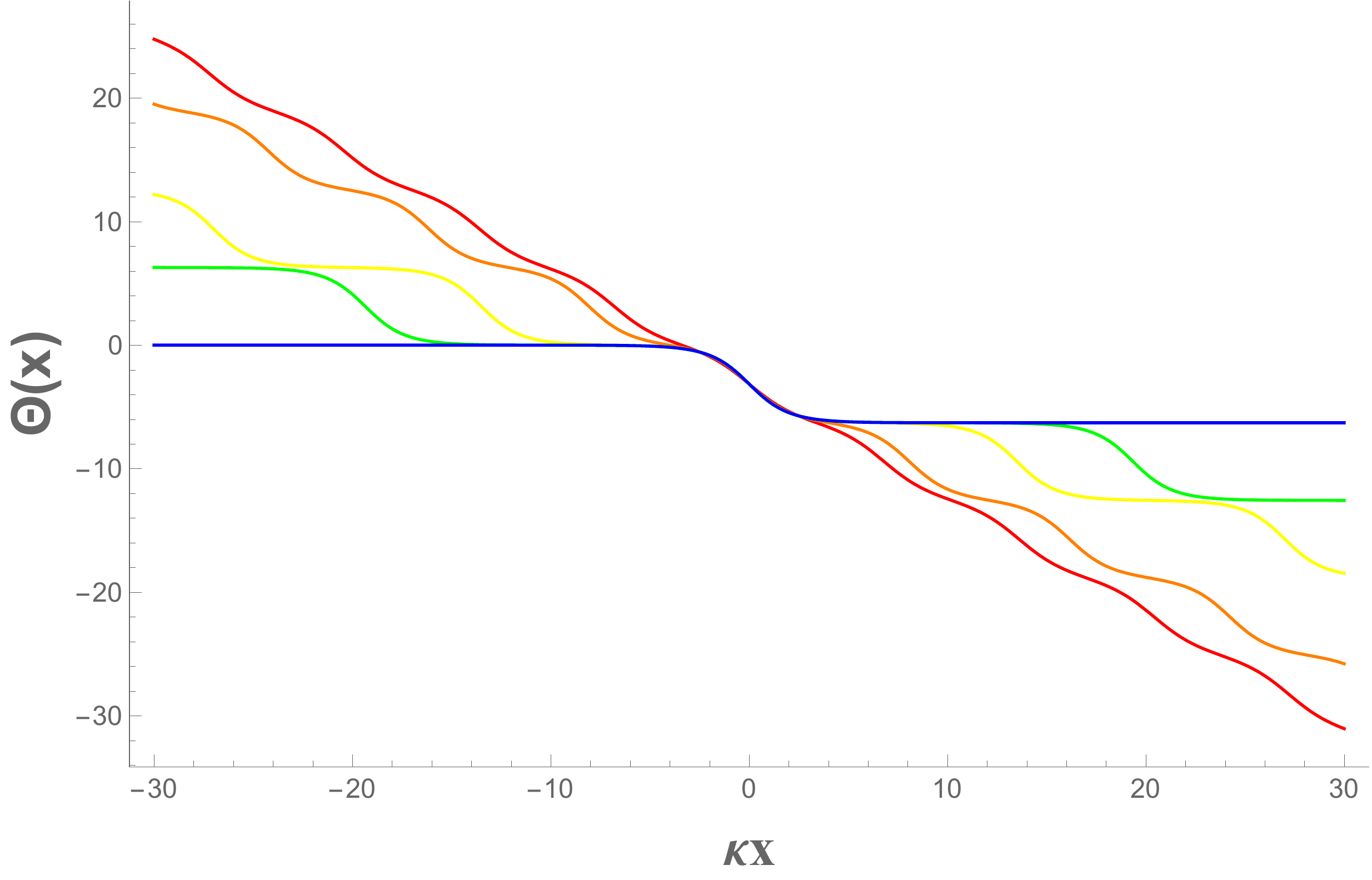}
\caption{These are plots of the profile $\Theta(x)$ against the dimensionless quantity $\kappa x$  for the lowest energy spiral configurations with $B(k)$ given by Eq.~\eqref{SG k minimum} , 
for several values of $k$: Red $k=\frac{9}{10}$, 
Orange $k=\frac{98}{100}$, Yellow $k=\frac{9998}{10000}$, Green 
$k=\frac{999998}{1000000}$, and Blue $k=1$. As the parameter $k$ 
increases $\Theta(x)$ tends towards a sine-Gordon domain wall 
configuration, achieving it for $k=1$.
}
\label{fig:external_field__spiral_state}
\end{center}
\end{figure}

As $C_0$ approaches the lowest allowed value $-U_{\rm min}=B$, 
the period $L$ becomes infinite, $f(C_0=-B;A=0,B)=0$, and the 
lowest energy spiral solution has the same average energy 
density as the ferromagnetic state. 
This phase boundary of spiral and ferromagnetic phases occurs at 
the $C_0\to-B$ limit of the minimum energy condition 
\eqref{SG k minimum}. 
%as \begin{equation}
%B=\frac{\pi^{2}}{16}\kappa^{2}, \label{critical mag field}
%\end{equation}because $E(k)\to 1$ as $k\to 1$. 
Taking into account the different unit conventions used 
when writing down the  effective energy density 
\eqref{eq:energy-density-solution}, our result on the critical 
magnetic field \eqref{SG k minimum}%{critical mag field} 
agrees with the previous 
result for $A=0$ \cite{BH}. 
 In Fig.~\ref{fig:external_field__spiral_state}, the profile 
function $\Theta(x)$ given in Eq.~\eqref{eq: solution_ellipticFunc_A=0} 
is plotted for several values of $k$ near 
the phase transition $B=\frac{\pi^{2}\kappa^2}{16}$.

As can be seen in Fig.~\ref{fig:external_field__spiral_state}, 
the domain wall solution connects $\Theta=0$ at $x=-\infty$ 
to $\Theta=-2\pi$ at $x=\infty$. 
The infinite period limit of one period of a spiral solution contains a
single domain wall in this case.


%\bibliographystyle{unsrt}
%\bibliography{calum-bib}
\end{document}